\documentclass[12pt,a4paper]{article}
\usepackage{ifthen} % for conditional statements
\newboolean{pdflatex}
\setboolean{pdflatex}{true} % False for eps figures 

\newboolean{articletitles}
\setboolean{articletitles}{true} % False removes titles in references

\newboolean{uprightparticles}
\setboolean{uprightparticles}{false} %True for upright particle symbols

\newboolean{inbibliography}
\setboolean{inbibliography}{false} %True once you enter the bibliography

\usepackage{microtype}
\usepackage{lineno}  % for line numbering during review
\usepackage{xspace} % To avoid problems with missing or double spaces after
\usepackage{caption} %these three command get the figure and table captions automatically small

\usepackage{graphicx}  % to include figures (can also use other packages)
\usepackage{color}
\usepackage{colortbl}
\graphicspath{{./figs/}} % Make Latex search fig subdir for figures

\usepackage{amsmath} % Adds a large collection of math symbols
\usepackage{amssymb}
\usepackage{amsfonts}
\usepackage{upgreek} % Adds in support for greek letters in roman typeset

\newcommand*\patchAmsMathEnvironmentForLineno[1]{%
\expandafter\let\csname old#1\expandafter\endcsname\csname #1\endcsname
\expandafter\let\csname oldend#1\expandafter\endcsname\csname
end#1\endcsname
 \renewenvironment{#1}%
   {\linenomath\csname old#1\endcsname}%
   {\csname oldend#1\endcsname\endlinenomath}%
}
\newcommand*\patchBothAmsMathEnvironmentsForLineno[1]{%
  \patchAmsMathEnvironmentForLineno{#1}%
  \patchAmsMathEnvironmentForLineno{#1*}%
}
\AtBeginDocument{%
\patchBothAmsMathEnvironmentsForLineno{equation}%
\patchBothAmsMathEnvironmentsForLineno{align}%
\patchBothAmsMathEnvironmentsForLineno{flalign}%
\patchBothAmsMathEnvironmentsForLineno{alignat}%
\patchBothAmsMathEnvironmentsForLineno{gather}%
\patchBothAmsMathEnvironmentsForLineno{multline}%
\patchBothAmsMathEnvironmentsForLineno{eqnarray}%
}

\usepackage{hyperref}    % Hyperlinks in references
\usepackage[all]{hypcap} % Internal hyperlinks to floats.

\def\MagUp {\mbox{\em Mag\kern -0.05em Up}\xspace}

\ifthenelse{\boolean{uprightparticles}}%
{

 \def\PDelta      {\ensuremath{\Delta}\xspace}                 
 \def\PXi      {\ensuremath{\Xi}\xspace}                 
 \def\PLambda      {\ensuremath{\Lambda}\xspace}                 
 \def\PSigma      {\ensuremath{\Sigma}\xspace}                 
 \def\POmega      {\ensuremath{\Omega}\xspace}                 
 \def\PUpsilon      {\ensuremath{\Upsilon}\xspace}                 
 
 \def\PB      {\ensuremath{\mathrm{B}}\xspace}                 
                  
 \def\PD      {\ensuremath{\mathrm{D}}\xspace}

 \def\PK      {\ensuremath{\mathrm{K}}\xspace}

 \def\Pi      {\ensuremath{\mathrm{i}}\xspace}

}
{

 \mathchardef\PDelta="7101
 \mathchardef\PXi="7104
 \mathchardef\PLambda="7103
 \mathchardef\PSigma="7106
 \mathchardef\POmega="710A
 \mathchardef\PUpsilon="7107
                  
 \def\PB      {\ensuremath{B}\xspace}                 
                  
 \def\PD      {\ensuremath{D}\xspace}

 \def\PK      {\ensuremath{K}\xspace}

 \def\Pi      {\ensuremath{i}\xspace}

}

\makeatletter
\ifcase \@ptsize \relax% 10pt
  \newcommand{\miniscule}{\@setfontsize\miniscule{4}{5}}% \tiny: 5/6
\or% 11pt
  \newcommand{\miniscule}{\@setfontsize\miniscule{5}{6}}% \tiny: 6/7
\or% 12pt
  \newcommand{\miniscule}{\@setfontsize\miniscule{5}{6}}% \tiny: 6/7
\fi
\makeatother

\DeclareRobustCommand{\optbar}[1]{\shortstack{{\miniscule (\rule[.5ex]{1.25em}{.18mm})}
  \\ [-.7ex] $#1$}}

  \def\Kbar    {{\kern 0.2em\overline{\kern -0.2em \PK}{}}\xspace}

\def\KorKbar    {\kern 0.18em\optbar{\kern -0.18em K}{}\xspace}

  \def\Dbar    {{\kern 0.2em\overline{\kern -0.2em \PD}{}}\xspace}

\def\DorDbar    {\kern 0.18em\optbar{\kern -0.18em D}{}\xspace}

\def\Bbar    {{\ensuremath{\kern 0.18em\overline{\kern -0.18em \PB}{}}}\xspace}

\def\BorBbar    {\kern 0.18em\optbar{\kern -0.18em B}{}\xspace}

  \def\Y#1S{\ensuremath{\PUpsilon{(#1S)}}\xspace}% no space before {...}!

\def\Lbar        {{\ensuremath{\kern 0.1em\overline{\kern -0.1em\PLambda}}}\xspace}
\def\LorLbar    {\kern 0.18em\optbar{\kern -0.18em \PLambda}{}\xspace}

\def\AT#1     {\ensuremath{A_{\mathrm{T}}^{#1}}\xspace}           % 2

\def\C#1      {\ensuremath{\mathcal{C}_{#1}}\xspace}                       % 9
\def\Cp#1     {\ensuremath{\mathcal{C}_{#1}^{'}}\xspace}                    % 7
\def\Ceff#1   {\ensuremath{\mathcal{C}_{#1}^{\mathrm{(eff)}}}\xspace}        % 9  
\def\Cpeff#1  {\ensuremath{\mathcal{C}_{#1}^{'\mathrm{(eff)}}}\xspace}       % 7
\def\Ope#1    {\ensuremath{\mathcal{O}_{#1}}\xspace}                       % 2
\def\Opep#1   {\ensuremath{\mathcal{O}_{#1}^{'}}\xspace}                    % 7

\newcommand{\tev}{\ifthenelse{\boolean{inbibliography}}{\ensuremath{~T\kern -0.05em eV}\xspace}{\ensuremath{\mathrm{\,Te\kern -0.1em V}}}\xspace}
\newcommand{\gev}{\ensuremath{\mathrm{\,Ge\kern -0.1em V}}\xspace}
\newcommand{\mev}{\ensuremath{\mathrm{\,Me\kern -0.1em V}}\xspace}
\newcommand{\kev}{\ensuremath{\mathrm{\,ke\kern -0.1em V}}\xspace}
\newcommand{\ev}{\ensuremath{\mathrm{\,e\kern -0.1em V}}\xspace}
\newcommand{\gevc}{\ensuremath{{\mathrm{\,Ge\kern -0.1em V\!/}c}}\xspace}
\newcommand{\mevc}{\ensuremath{{\mathrm{\,Me\kern -0.1em V\!/}c}}\xspace}
\newcommand{\gevcc}{\ensuremath{{\mathrm{\,Ge\kern -0.1em V\!/}c^2}}\xspace}
\newcommand{\gevgevcccc}{\ensuremath{{\mathrm{\,Ge\kern -0.1em V^2\!/}c^4}}\xspace}
\newcommand{\mevcc}{\ensuremath{{\mathrm{\,Me\kern -0.1em V\!/}c^2}}\xspace}

\def\mum  {\ensuremath{{\,\upmu\rm m}}\xspace}

\def\gsim{{~\raise.15em\hbox{$>$}\kern-.85em
          \lower.35em\hbox{$\sim$}~}\xspace}
\def\lsim{{~\raise.15em\hbox{$<$}\kern-.85em
          \lower.35em\hbox{$\sim$}~}\xspace}

\def\tell1  {TELL1\xspace}
\def\ukl1   {UKL1\xspace}

\usepackage{cite} % Allows for ranges in citations
\usepackage{mciteplus}

\usepackage{longtable} % only for template; not usually to be used in PAPERs

\begin{document}

%%%%%%%%%%%%%%%%%%%%%%%%%
%%%%% Title     %%%%%%%%%
%%%%%%%%%%%%%%%%%%%%%%%%%
\renewcommand{\thefootnote}{\fnsymbol{footnote}}
\setcounter{footnote}{1}

% %%%%%%% CHOOSE TITLE PAGE--------
%\onecolumn
%\input{title-LHCb-ANA}
%\input{title-LHCb-CONF}
%\input{title-LHCb-PAPER}

% $Id: title-LHCb-PAPER.tex 67452 2015-02-10 11:22:35Z roldeman $
% ===============================================================================
% Purpose: LHCb-PAPER journal paper title page template
% Author: 
% Created on: 2010-09-25
% ===============================================================================

%%%%%%%%%%%%%%%%%%%%%%%%%
%%%%%  TITLE PAGE  %%%%%%
%%%%%%%%%%%%%%%%%%%%%%%%%
\begin{titlepage}
\pagenumbering{roman}

% Header ---------------------------------------------------
\vspace*{-1.5cm}
\centerline{\large EUROPEAN ORGANIZATION FOR NUCLEAR RESEARCH (CERN)}
\vspace*{1.5cm}
\hspace*{-0.5cm}
\begin{tabular*}{\linewidth}{lc@{\extracolsep{\fill}}r}
\ifthenelse{\boolean{pdflatex}}% Logo format choice
{\vspace*{-2.7cm}\mbox{\!\!\!\includegraphics[width=.14\textwidth]{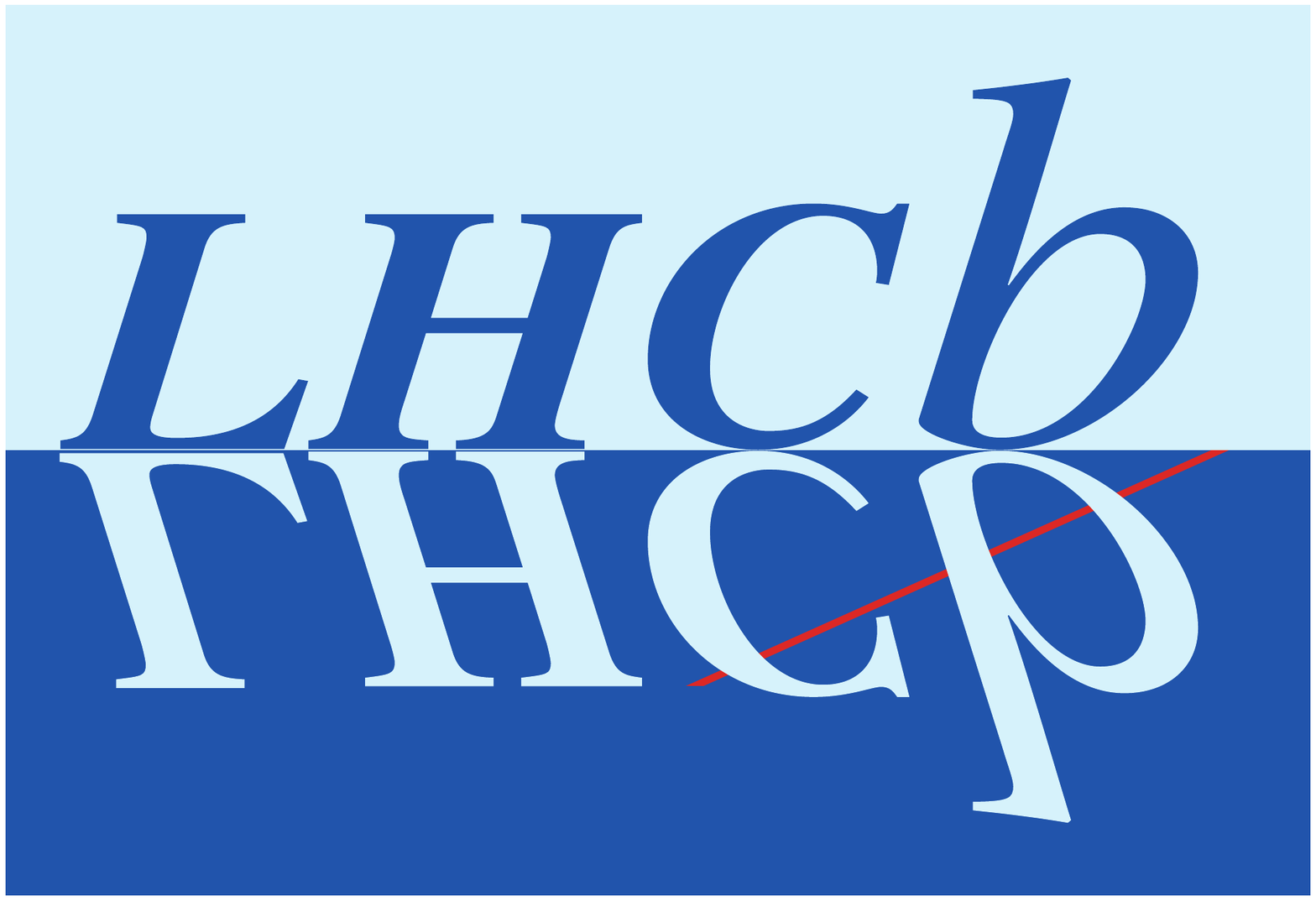}} & &}%
{\vspace*{-1.2cm}\mbox{\!\!\!\includegraphics[width=.12\textwidth]{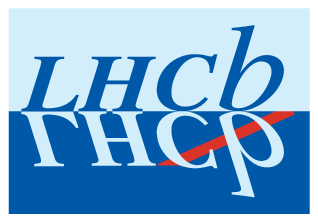}} & &}%
\\
 & & LHCb-PUB-2015-015 \\  % ID 
 & & Oct 9, 2015 \\ % Date - Can also hardwire e.g.: 23 March 2010
 & & \\
% not in paper \hline
\end{tabular*}

\vspace*{0.25cm}

% Title --------------------------------------------------
{\bf\boldmath\huge
\begin{center}
  Testbeam studies of pre-prototype silicon strip sensors for the LHCb UT upgrade project
\end{center}
}

\vspace*{2.0cm}

% Authors -------------------------------------------------
\begin{center}
A.~Abba$^1$, M.~Artuso$^2$, S.~Blusk$^{2,6}$, T.~Britton$^2$, A.~Davis$^3$, A.~Dendek$^4$, B.~Dey$^5$, S.~Ely$^2$,  T.~Evans$^{7}$, J.~Fu$^1$, P.~Gandini$^2$, F.~Lionetto$^5$, P.~Manning$^2$, B.~Meadows$^3$, R. Mountain$^2$, N.~Neri$^1$, M.~Petruzzo$^1$, M.~Pikies$^4$, T.~Skwarnicki$^2$, T.~Szumlak$^4$, J.~C.~Wang$^2$   
\bigskip\\
{\it\footnotesize
$ ^1$Instituto Nazionale di Fisica Nucleare - Sezione di Milano, Italy \\
$ ^2$Syracuse University, Syracuse, NY USA \\
$ ^3$University of Cincinnati, Cincinnati, OH USA \\
$ ^4$AGH - University of Science and Technology, Faculty of Physics and Applied Computer Science, Krak\'{o}w, Poland\\
$ ^5$Physik-Institut, Universit\"{a}t Z\"{u}rich, Z\"{u}rich, Switzerland \\
$ ^6$European Organization for Nuclear Research (CERN), Geneva, Switzerland\\
$ ^{7}$Department of Physics, University of Oxford, Oxford, United Kingdom
}
\end{center}

\vspace{\fill}

% Abstract -----------------------------------------------
\begin{abstract}
  \noindent
  The LHCb experiment is preparing for a major upgrade in 2018-2019. One of the key
  components in the upgrade is a new silicon tracker situated upstream of the analysis magnet of the experiment.
  The Upstream Tracker (UT) will consist of four planes of silicon strip detectors, with each plane
  covering an area of about 2~m$^2$. An important consideration of these detectors is their performance 
  after they have been exposed to a large radiation dose. In this article we present test beam results 
  of pre-prototype n-in-p and p-in-n sensors that have been irradiated with fluences up to $4.0\times10^{14}~n_{\rm eq}$/cm$^2$.
\end{abstract}

\vspace*{1.0cm}

\begin{center}
  Published in Nucl. Instrum. Meth. A.
\end{center}

\vspace{\fill}

{\footnotesize 
\centerline{\copyright~CERN on behalf of the LHCb collaboration, licence \href{http://creativecommons.org/licenses/by/4.0/}{CC-BY-4.0}.}}
\vspace*{2mm}

\end{titlepage}

%%%%%%%%%%%%%%%%%%%%%%%%%%%%%%%%
%%%%%  EOD OF TITLE PAGE  %%%%%%
%%%%%%%%%%%%%%%%%%%%%%%%%%%%%%%%

%  empty page follows the title page ----
\newpage
\setcounter{page}{2}
\mbox{~}
%\newpage

\cleardoublepage

%\twocolumn
% %%%%%%%%%%%%% ---------

\renewcommand{\thefootnote}{\arabic{footnote}}
\setcounter{footnote}{0}

%%%%%%%%%%%%%%%%%%%%%%%%%%%%%%%%
%%%%%  Table of Content   %%%%%%
%%%%%%%%%%%%%%%%%%%%%%%%%%%%%%%%
%%%% Uncomment next 2 lines if desired
%\tableofcontents
%\cleardoublepage

%%%%%%%%%%%%%%%%%%%%%%%%%
%%%%% Main text %%%%%%%%%
%%%%%%%%%%%%%%%%%%%%%%%%%

\pagestyle{plain} % restore page numbers for the main text
\setcounter{page}{1}
\pagenumbering{arabic}

%% Uncomment during review phase. 
%% Comment before a final submission.
%\linenumbers

% You can include short sections directly in the main tex file.
% However, for larger papers it is desirable to split the text into
% several semiautonomous files, which can be revised independently.
% This is especially useful when developing a document in
% collaboration with several people, since then different parts can be
% edited independently.  This type of file organization is shown here.
% 

%%%%\input{introduction}
\section{Introduction}
\label{sec:Introduction}
The Upstream Tracker (UT) detector is a key part of the LHCb Upgrade, replacing the current TT tracking 
stations~\cite{LHCb-TDR-008,LHCb-TDR-009}. The UT improves over the current TT in that it
(i) eliminates all gaps within the detector acceptance, (ii) has largely improved granularity to cope with the
higher instantaneous luminosity expected in the LHCb upgrade, and (iii) improves the coverage close to
the beam pipe by employing a circular cutout to match the beam pipe profile.
The UT, like the TT, is situated just in front of LHCb's dipole analysis magnet. In this position, it
provides a crucial link between segments reconstructed in the upgraded vertex detector~\cite{LHCb-TDR-013}
and the tracking chambers downstream~\cite{LHCb-TDR-015} of the LHCb magnet.
It provides a factor of 3 improvement in speed for the
tracking in the the fully-software-based trigger~\cite{LHCb-TDR-016}, reduces the rate of fake tracks being
formed by a factor of 2-3, and improves the momentum resolution by about 25\% relative to tracks not
using UT hits. The factor of 3 in speed is enabled by enabling a very fast estimate of the momentum
of charged particles, which can then be used to reduce the size of the hit search windows in the downstream
tracking stations. Due to the increased speed of the trigger and the higher purity of tracks considered,
larger data sets with better signal-to-background can be acquired.
\begin{figure}[b]
\centering
\vspace{-0.5in}
\includegraphics[width=1.0\textwidth]{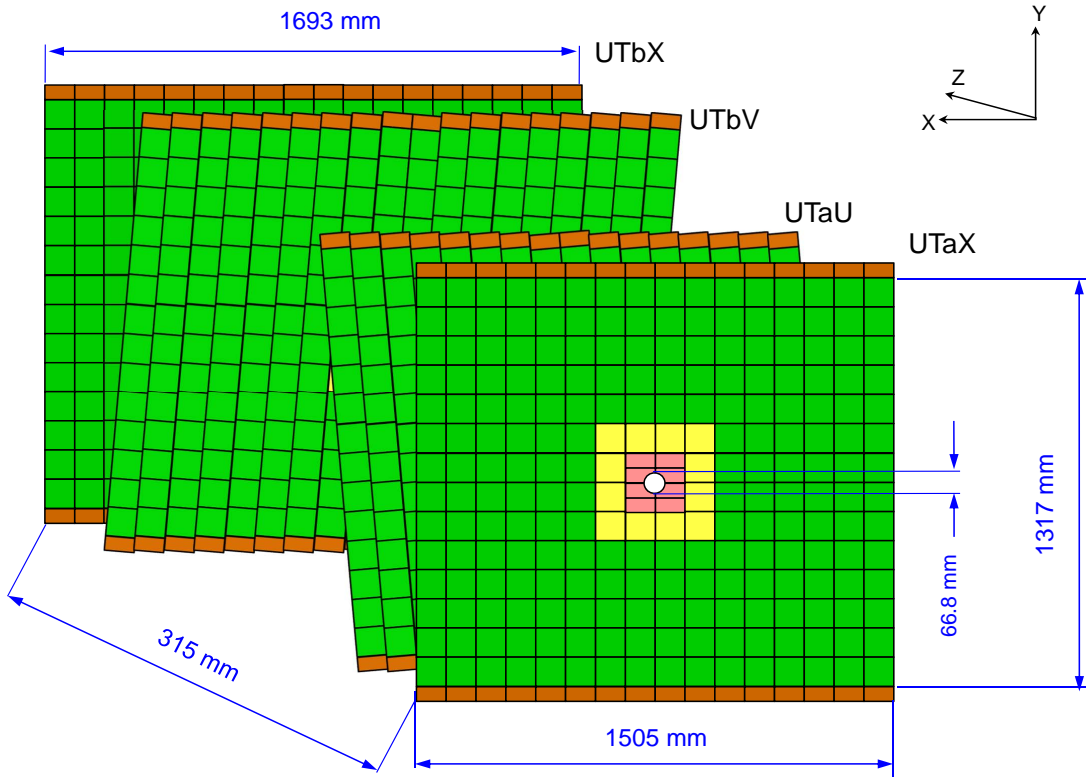}
\caption{\small{Cartoon showing the layout and dimensions of the four UT planes.}}
\label{fig:ut_cartoon}
\end{figure}

The UT detector consists of four silicon planes,
each about 1.53 m in width and 1.34 m in height. Each plane is composed of 1.5 m long {\it staves} that are tiled with
$\sim$10 cm~$\times$~10~cm silicon wafers. Consecutive wafers are mounted on opposite sides of the stave to ensure no gaps
along the height, and adjacent staves are also overlapped to ensure no gaps in the horizontal direction.
The majority of the detector area utilizes sensors with an approximate pitch of $190~\mu$m, however the inner region 
features sensors with half the pitch ($95~\mu$m) to cope with higher occupancy. 
%A cartoon of the system is shown in Fig.~\ref{fig:ut_cartoon}.
Both n-in-p and p-in-n type sensors are being considered for the outer region, but for the inner region, only
n-in-p are being considered due to better radiation hardness. For the innermost region of the UT, the
largest fluence expected, with a safety factor of about two, is about $5\times10^{14}~n_{\rm eq}$/cm$^2$. Outside this region,
the fluence is not expected to exceed about $2\times10^{13}~n_{\rm eq}$/cm$^2$.

The primary goals of this test beam were to quantify the performance of several pre-prototype n-in-p mini-sensors
from Micron Semiconductor, Ltd~\cite{micron}, after a high radiation fluence, and compare to corresponding results from
similar unirradiated detectors. The properties investigated include, but are not limited to:
\begin{itemize}
\item Landau distributions as a function of bias voltage;
\item Cluster size versus bias voltage and angle of incidence;
\item Charge collected versus interstrip position;
\item Resolution versus angle;
\item Characteristics of sensors near the quarter-circle region (these emulate the sensors surrounding the beam pipe).
\end{itemize}
One p-in-n mini-sensor irradiated to the maximum level expected in the outer region of the UT was also tested, but its
study was not a primary focus of the test beam results presented here.

\section{Experimental setup}
\label{sec:expSetup}

The test beam discussed in this article was conducted in October 2014 at the SPS at CERN. 
The beam consisted of positively charged hadrons with momentum of 180~\gevc. The beam was delivered in spills at rate of about 4 spills/minute, 
with each spill lasting about 4 seconds. For most of the data taking, the beam size was collimated down to about 0.5 cm in diameter
and each spill provided a particle rate of order 1 MHz.

\subsection{Telescope description}
\begin{figure}[tb]
\centering
\includegraphics[width=1.0\textwidth]{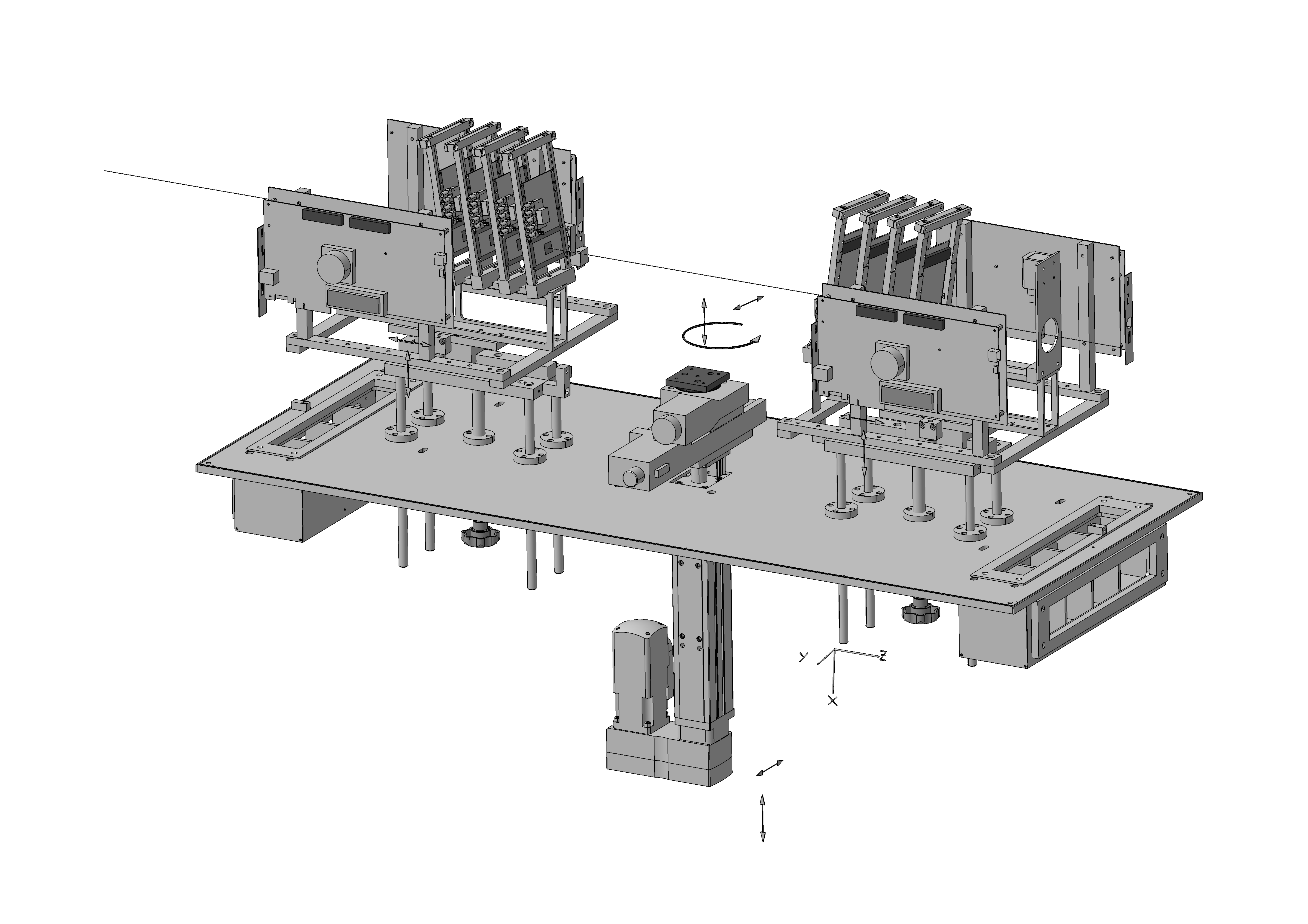}
\caption{\small{Cartoon showing the TimePix3 telescope~\cite{Akiba:2013yxa} used to study the UT sensors.}}
\label{fig:tp3_cartoon}
\end{figure}

The pre-prototype UT sensors, or detectors under test (DUT), were studied using the TimePix3 (TP3)-based telescope~\cite{Akiba:2013yxa}, composed of 8 pixel planes. 
Each pixel plane is about $1.4~\mathrm{cm}\times 1.4~\mathrm{cm}$ and has a pixel size of $55~\mu\mathrm{m} \times 55~\mu\mathrm{m}$. The planes
are tilted in order to provide more charge sharing, and thus better position resolution.
A cartoon of the telescope layout is shown in Fig.~\ref{fig:tp3_cartoon}.
With the high momentum beam of 180\gevc, the reconstructed tracks provide excellent pointing resolution of about 2~$\mu$m at the DUT.
The telescope readout does not require an external trigger: hits are recorded continuously once a run is started. 
For each pixel hit, both position and a time-stamp with 1.56~ns precision is recorded. Tracks are then formed (offline) by combining hits
that have compatible time values.

\subsection{Detectors under test} 
\begin{figure}[tb]
\centering
\includegraphics[width=1.0\textwidth]{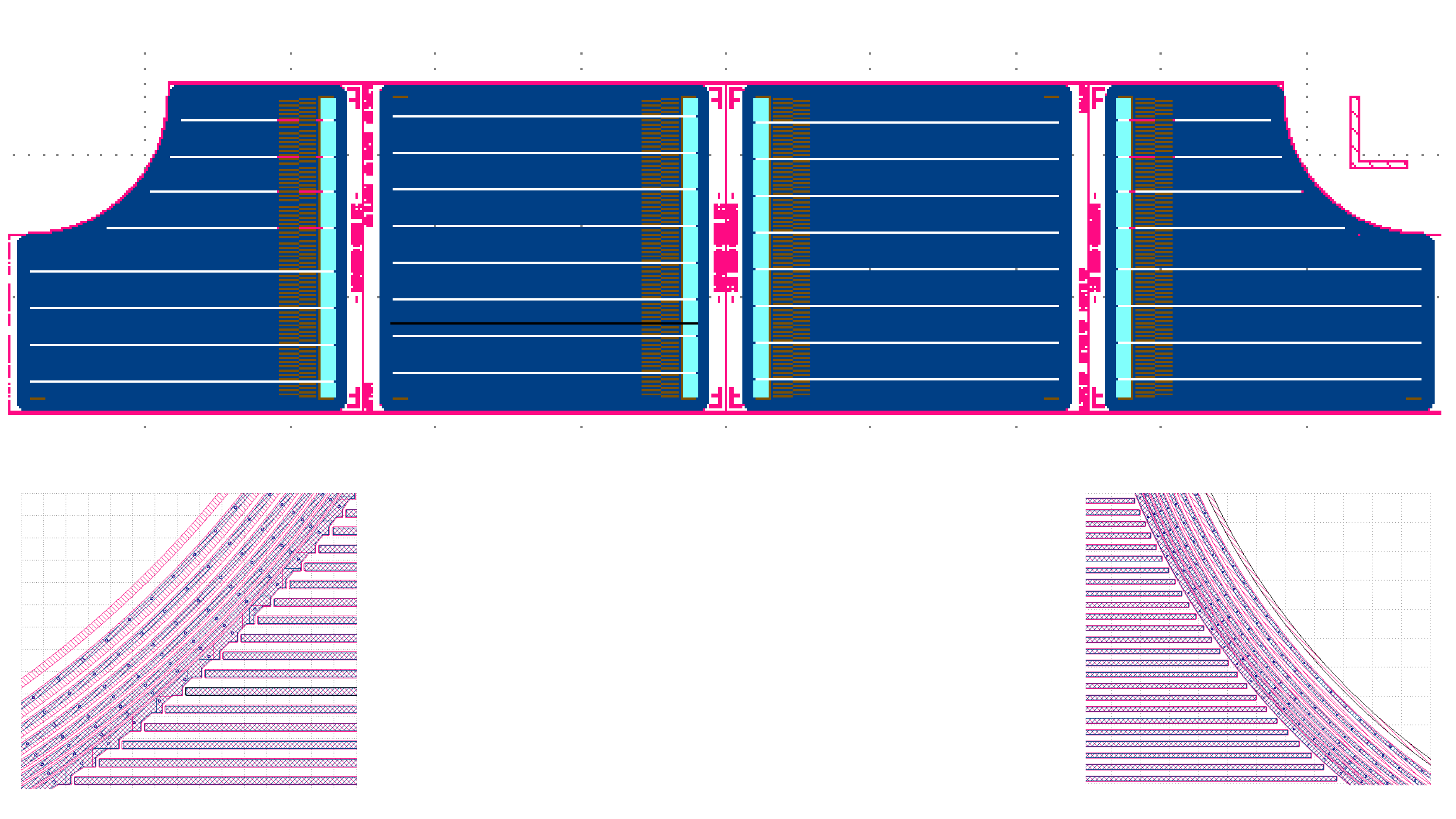}
\caption{\small{Schematic of the mini-sensors tested during the October 2014 test beam. The two sensors in the middle are
full mini-sensors, while the ones on the left and right have a quarter-circle cut out. The difference between
the left-most (MBP1) and right-most sensors (MBP2) is in the guard-ring structure, as shown just below them. }}
\label{fig:sensors}
\end{figure}

The {\it mini-sensor} pre-prototypes tested in the October 2014 test beam were obtained from Micron Semiconductor, Ltd~\cite{micron}. 
One of the sensors tested was a p-in-n, and six were n-in-p. The resistivities of the sensors, as determined from capacitance versus voltage
measurements, were about 0.9 $k\Omega$~cm for the p-in-n sensor and about 2.8 $k\Omega$~cm for the n-in-p.
Each sensor was 1.115 cm x 1.125 cm in size with a nominal thickness of 250~$\mu$m, and had 128 strips with a strip pitch of 
80~$\mu$m and a strip (implant) width of 30~$\mu$m.  Prior to irradiation, all the of sensors had a depletion voltage of about 180~V.
A schematic of the mini-sensors is shown in Fig.~\ref{fig:sensors}. In this schematic, the strips run horizontally. 
Of the six n-in-p sensors tested, three had the strips terminated such that they form a quarter-circle inactive region
of the sensor. Two different guard-ring structures were implemented, as shown in Fig.~\ref{fig:sensors}. 
More details of the sensors under test are shown in Table~\ref{tab:sensor-details}. The MBP1 and MBP2 sensors 
are differentiated by the guard ring structures, and are shown on the left and right side of Fig.~\ref{fig:sensors},
respectively. The MBP1 (leftmost sensor in figure) differs from the MBP2 sensor (rightmost) in that it implements a stepped structure
along the innermost guard ring to maintain an equal separation between the edge of the strip and the innermost guard ring, 
more like a conventional rectangular-shaped detector. 

ix of the seven sensors were irradiated at the Massachusetts General Hospital (MGH) proton irradiation facility~\cite{MGH-PTC} 
in June 2014, using protons of kinetic energy equal to 226 $\mev$ and fluences ranging from 
$0.27\times10^{14}~n_{\rm eq}$/cm$^2$ to $4.0\times10^{14}~n_{\rm eq}$/cm$^2$. 
\begin{table*}[tb]
\begin{center}
\caption{\small{Specific details of the senors tested in the October 2014 test beam. Here, 
$\rho$ refers to the resistivity of the silicon and $N_{e,h}$ indicates either the electron or hole concentration.
For sensor 3091-10-MBP2, CV measurements were not available.}}
\begin{tabular}{lccccc}
\hline\hline
Sensor     &      Type    & Fluence                         & $N_{e,h}$              & $\rho$       & Other \\
  ID       &              & ($10^{14}~n_{\rm eq}$/cm$^2$)  & ($10^{12}$ cm$^{-3}$) & ($k\Omega$~cm) & spec. \\
\hline
3092-1-MS2      & p-in-n  &    0.27  & 5.1 & 0.87   &  250~$\mu$m \\
3091-8-MS1      & n-in-p  &    1.0  & 5.0 & 2.79   & 242~$\mu$m, p-spray \\
3091-10-MBP2    & n-in-p  &    0.0  & n/a & n/a     & 238~$\mu$m, p-spray \\
3091-7-MS2      & n-in-p  &    4.0 & 5.0 & 2.76  & 254~$\mu$m, p-spray\\
3091-7-MS1      & n-in-p  &    1.8 & 4.9 & 2.85  & 254~$\mu$m, p-spray\\
3091-8-MBP2     & n-in-p  &    4.0 & 5.2 & 2.69  & 242~$\mu$m, p-spray\\
3091-8-MBP1     & n-in-p  &    4.0 & 5.0 & 2.79  & 242~$\mu$m, p-spray\\
\hline\hline
\end{tabular}
\label{tab:sensor-details}
\end{center}
\end{table*}
Between the time of the irradiation in June 2014, and the test beam in October 2014, the sensors were kept
in a freezer, at a temperature below -10~$^\circ$C. The sensors were warmed up to room temperature for
no more than 7 days to transport the sensors and for wirebonding.

The DUT readout for this testbeam was based on the Alibava DAQ 
system~\cite{Alibava,Bernabeu:2013qta,MarcoHernandez:2011zz,MarcoHernandez:2009zza}, 
which uses Beetle chips~\cite{Beetle, LoechnerThesis} as the front end ASIC. 
The main components of the Alibava system are a 
detector board to which sensors are mounted, a {\it daughterboard} that includes two Beetle chips (128 channels each),
and a {\it motherboard} that manages the data flow to/from the Beetle chips and to/from the data acquisition PC. 

\begin{figure}[tb]
\centering
\includegraphics[width=1.05\textwidth]{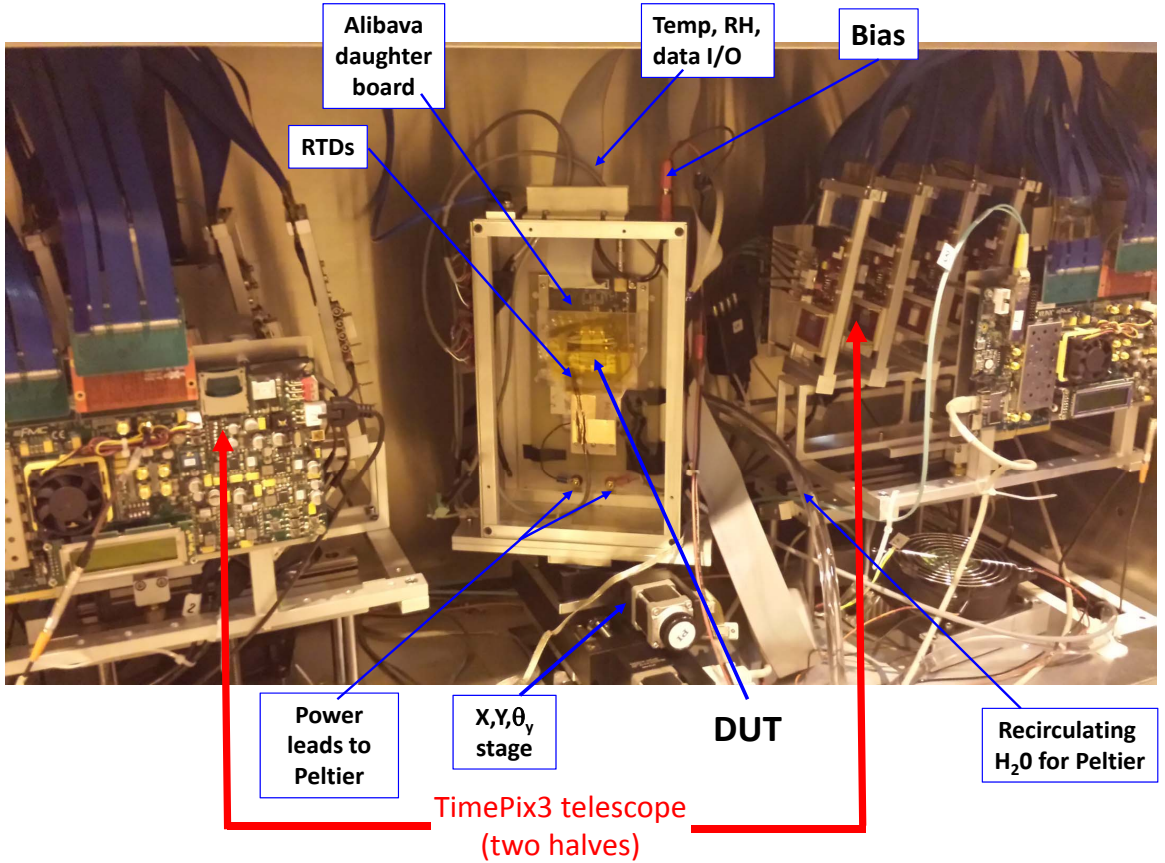}
\caption{\small{Photograph of the TimePix3 telescope and the DUT during installation at the SPS, together with a description of the
different components.}}
\label{fig:PhotoOfTelescope}
\end{figure}

The DUT was housed in an Aluminum box to provide both shielding and a light-tight environment. A Peltier device was used to
cool the sensor to about -13~$^{\rm o}$C, and a continuous nitrogen flow maintained the relative humidity close to zero.
The entire box was mounted on a stage that allowed for horizontal and vertical translations,
as well as rotations about the vertical axis to allow for studies of the detector performance versus particle incident angle.
The system was equipped with both temperature and relative humidity monitoring.
A photograph of the telescope with the UT DUT being installed is shown in Fig.~\ref{fig:PhotoOfTelescope}. 
Various components are indicated.

\subsection{Trigger and synchronization}

The trigger was formed from the coincidence of two scintillators, one just upstream and a second just downstream of the
TimePix3 telescope (see Fig.~\ref{fig:tp3_cartoon}) and a {\bf not-busy} 
signal from the Alibava motherboard. This trigger signal was fanned out to both the TimePix3 telescope and the Alibava motherboard. 
Because the data acquisition systems of the TimePix3 telescope and the DUT were independent of one another, the common trigger 
signal was used to synchronize the two systems. The signal sent to the Alibava system
initiated a readout of the DUT, while the signal sent to the TimePix3 system produced a timestamp in the TimePix3 event record.
This timestamp used the same clock used to time stamp the pixel hits, and in this way, pixel hits could be associated with
a specific trigger. Since the number of time stamps in the TimePix3 event record was identical to the number of triggers
sent to the Alibava system, the synchronization only required matching the first event in the Alibava system with the
pixel hits/tracks associated to the first trigger time stamp, and then the second event matched to the second trigger time stamp,
and so on. In this way, the beam tracks reconstructed in the TimePix3 system were properly matched to the corresponding 
hits produced in the DUT.

\section{Corrections and calibrations}
\label{sec:CalandCorr}

Before being able to quantify the performance of the sensors, various calibrations and corrections were applied.
These corrections included pedestal and common-mode noise suppression, a cross-talk correction, an
out-of-time correction, and are detailed below.

\subsection{Calibration scans}
Calibration scans, which were taken periodically during the test beam, showed that 1 ADC was equivalent to about 275~$e^-$.
The exact value varied by a few percent, depending on the sensor/readout board. Since the calibration circuit has its own
inherent parasitic capacitances, we estimate that there is an inherent uncertainty in the overall calibration not larger than
about 5\%. Charges presented throughout this article are given in terms of ADC counts, and must be multiplied by 275 to obtain
the corresponding charge in number of electrons. 

\subsection{Pedestal and common mode noise suppression} \label{sec:Ped}

Pedestals are calculated from datasets of $10^4-10^5$~events recorded during beam stops and are then subtracted from the
beam-on runs recorded under the same conditions. (The events are self-triggered through the Alibava motherboard).
Dead or noisy strips are excluded from all parts of the analysis.
The pedestal of a given Beetle channel is evaluated as the mean of the raw 
ADC counts of that channel over the whole pedestal run, after excluding anomalously large or small ADC values.
After subtracting the pedestals, the common mode noise is computed (event-by-event) as the average ADC value in a given chip, 
and is then subtracted from each channel. After the pedestal and common mode noise suppression, the
noise level is about 3.5 ADC counts, or about 1000 electrons.

\subsection{Cross-talk correction}
\begin{figure}[tb]
\vspace{-1.0in}
\centering
\includegraphics[width=1.0\textwidth]{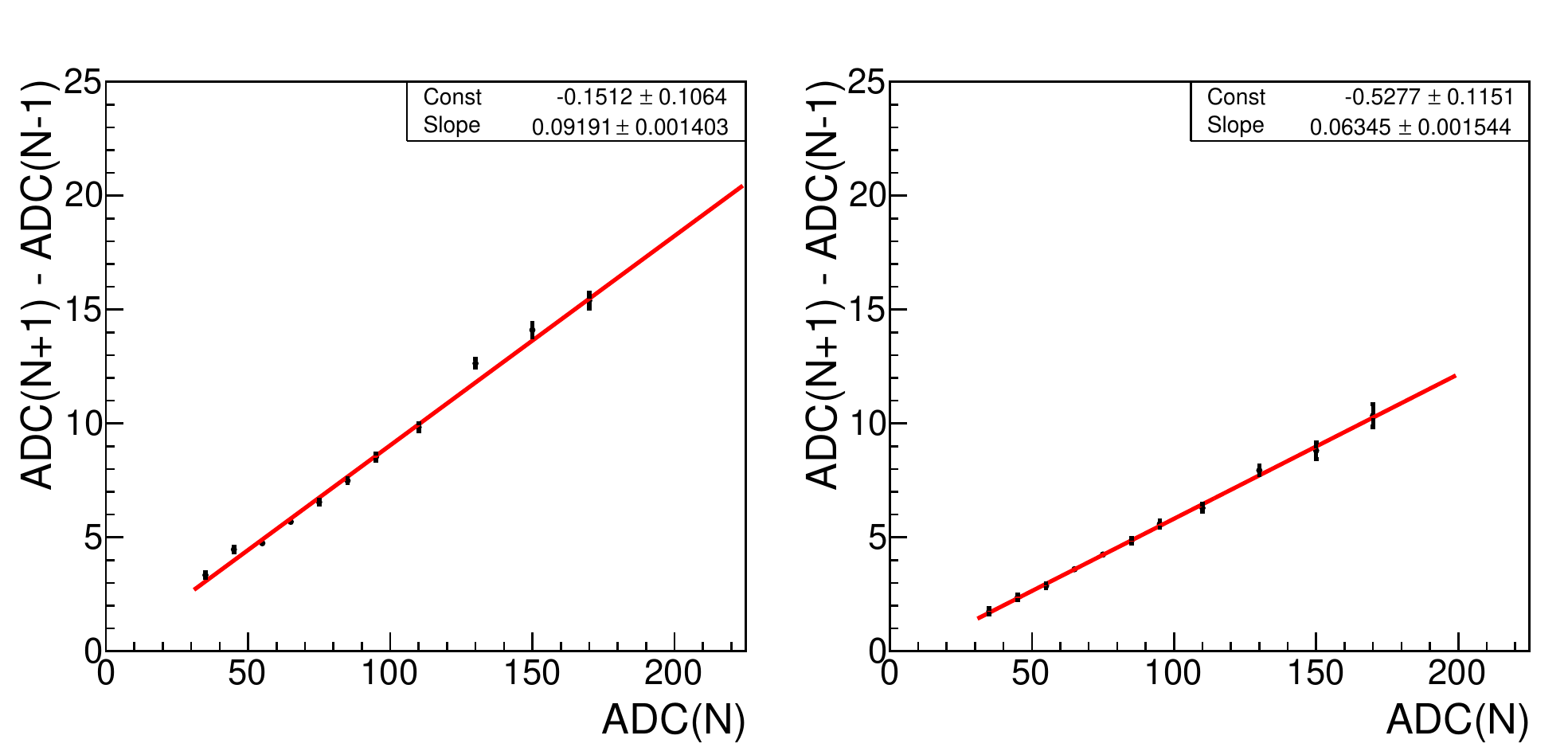}
\includegraphics[width=1.0\textwidth]{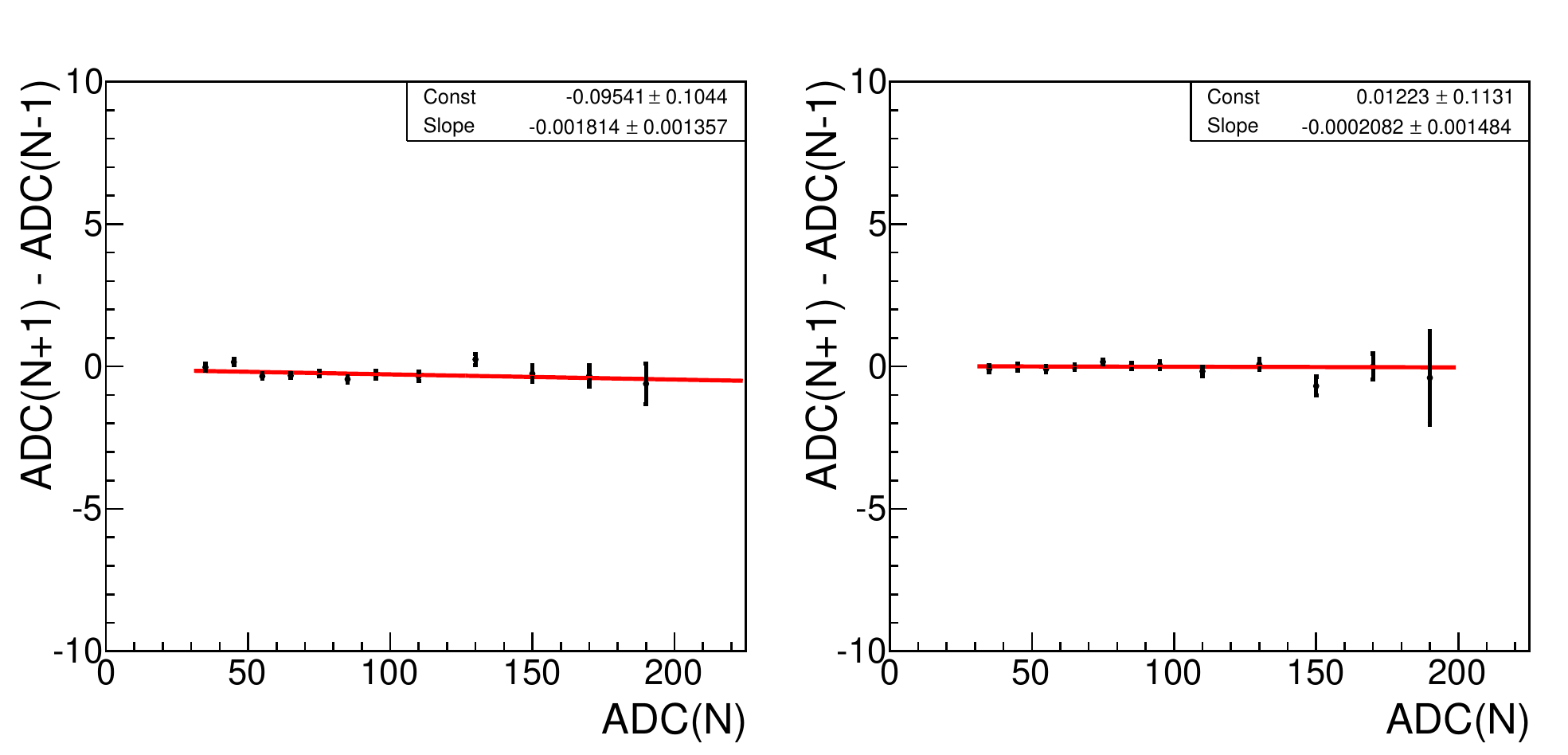}
\caption{\small{Mean of the difference in charge detected, $\langle$ADC(N+1)-ADC(N-1)$\rangle$ between the (N+1)$^{\rm th}$ and (N-1)$^{\rm th}$ strips, 
as a function of the charge seen in the N$^{\rm th}$ strip, for sensor 3092-1-MS2. The top pair of plots show the values for even and odd channels before the correction, 
the bottom pair after the correction. Plots with an even peak strip numbers are shown on the left, while plots with an odd peak strip number are shown on the right. The red lines are linear fits to the data.}}
\label{fig:CrossTalk}
\end{figure}

From analysis of the data, it was found that there was cross-talk between the N$^{\rm th}$ and the (N$+1)^{\rm th}$ channel.
The effect was easily seen by studying the charge asymmetry between the (N-1)$^{\rm th}$ and (N$+1)^{\rm th}$ channel about
the peak strip ($N$) in a cluster with tracks at normal incidence. Due to capacitative coupling some charge sharing
is expected, but it should not be asymmetric. The asymmetry is studied for even and odd-numbered peak strips separately.
The resulting asymmetries are shown in Fig.~\ref{fig:CrossTalk} (top) before the correction and (bottom) after the linear correction
for one of the DUTs. 
The even channels show a larger cross-talk than the odd channels. An empirical correction based on the linear fit shown
is applied, and the asymmetry after the correction is shown. This is believed to be due to a sub-optimal timing of the ADC sampling 
phase, which was a fixed setting in the firmware of the motherboard. The Beetle chip itself is known to have a small amount of
cross-talk~\cite{LoechnerThesis}, but the level is below 2\%; thus most of the effect is attributable to the readout.
All DUTs have similar odd-even cross-talk corrections.
The even/odd cross-talk correction was similar for all of the DUTs, but not identical. Therefore, each sensor had a unique cross-talk correction
to remove this bias.

\subsection{Out-of-time correction}
\label{sec:out_of_time_correction}
\begin{figure}[tb]
\centering
\includegraphics[width=0.48\textwidth]{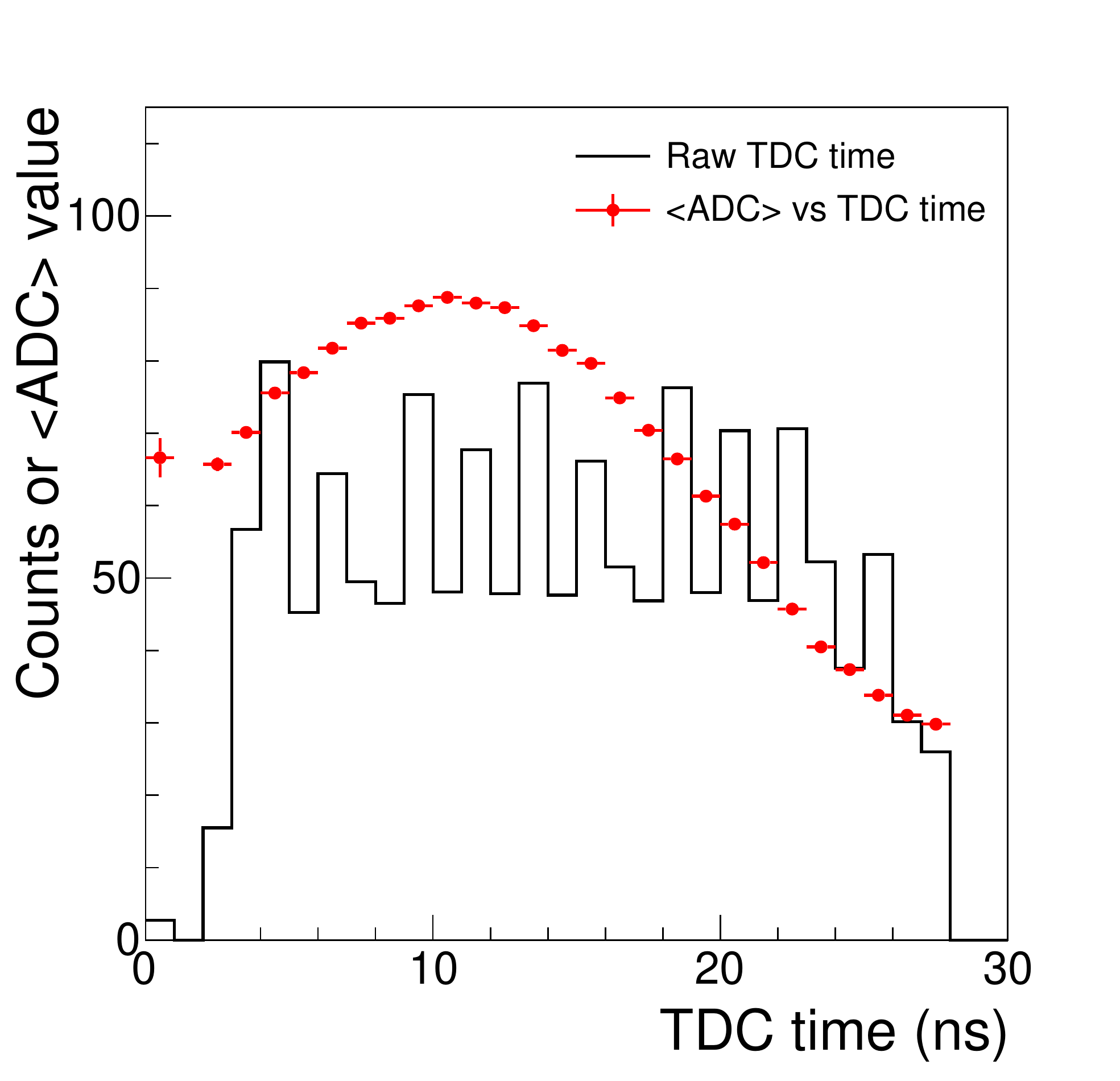}
\includegraphics[width=0.48\textwidth]{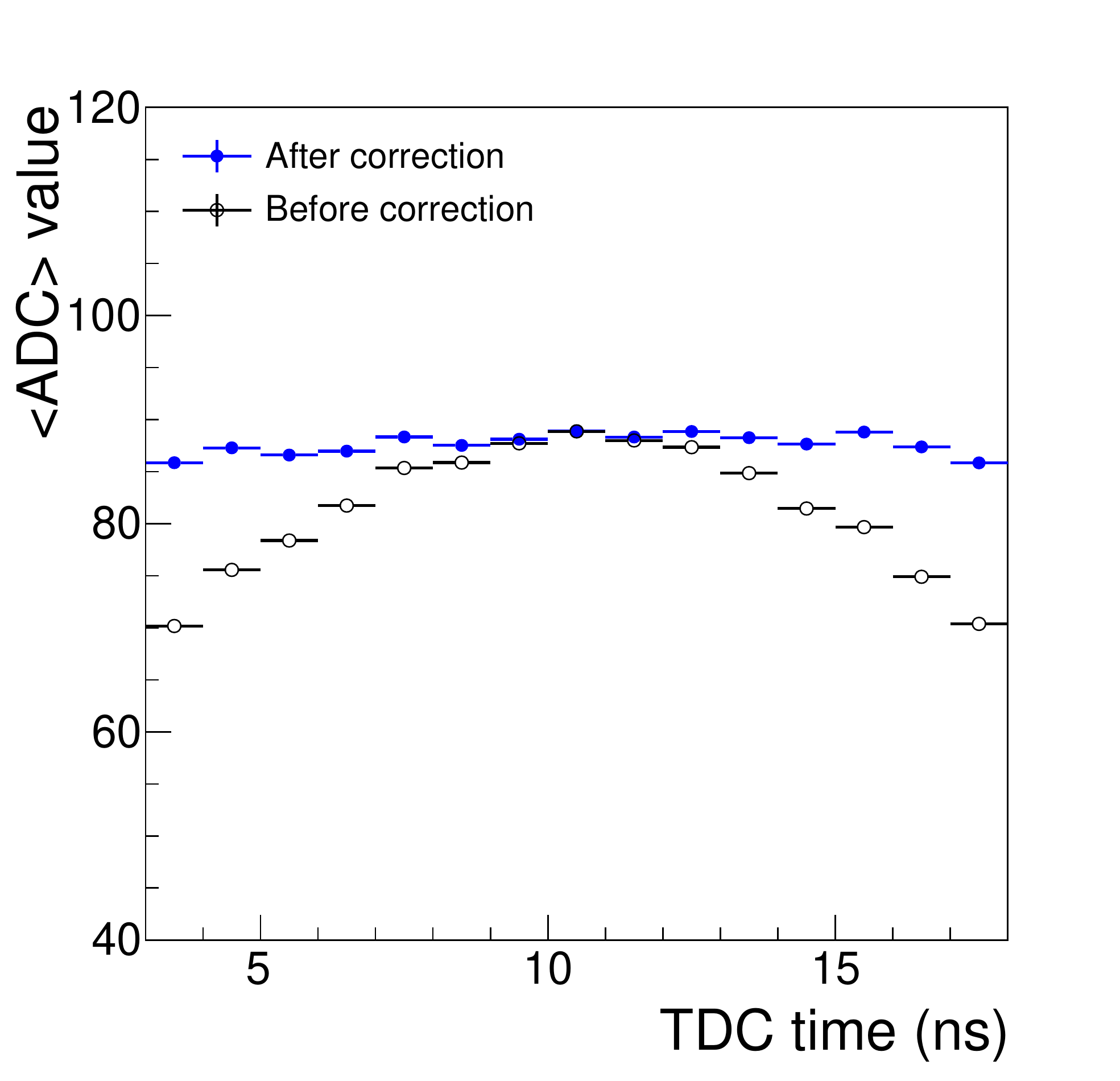}
\caption{\small{(Left) Raw TDC time spectrum, overlaid with the average ADC versus TDC time for DUT signal clusters. (Right)
Average ADC versus TDC time for DUT signal clusters before (black points) and after (blue points) the time correction.}}
\label{fig:TDCSpectraCorr}
\end{figure}

The Alibava system provides a 40 MHz clock to sample the signals from DUT in the front end of the Beetle chip. 
However, beam particles arrive asynchronously with respect to this clock. The Alibava system stores the TDC time 
between the signals from the scintillators and the edge of the 40 MHz clock. In the offine analysis, the sampled 
pulse heights are sorted with the TDC values. When timing in the system, we adjusted the latency such that the signal 
from a beam particle arrives at about 10 ns. That is, the maximum charge is collected from the Beetle when the TDC time is 10 ns. 
To make use of the data not precisely at the peak, we include and correct the measured charge for signals within 
$\pm$3.5 ns of the peak.
The correction is obtained by fitting the average ADC value for signal clusters as a function of the
TDC time to a Gaussian function. In this limited range, the correction is no more than about $5-6\%$. 
Figure~\ref{fig:TDCSpectraCorr}\,(left) shows the raw time spectrum of all triggered events, overlaid with the average
ADC for signal clusters as a function of the TDC time. The right panel shows the average ADC versus time before and after the correction,
where only the region used in the fit is shown.
%\clearpage

\subsection{Alignment}

\begin{figure}[tb]
\centering
\includegraphics[width=1.0\textwidth]{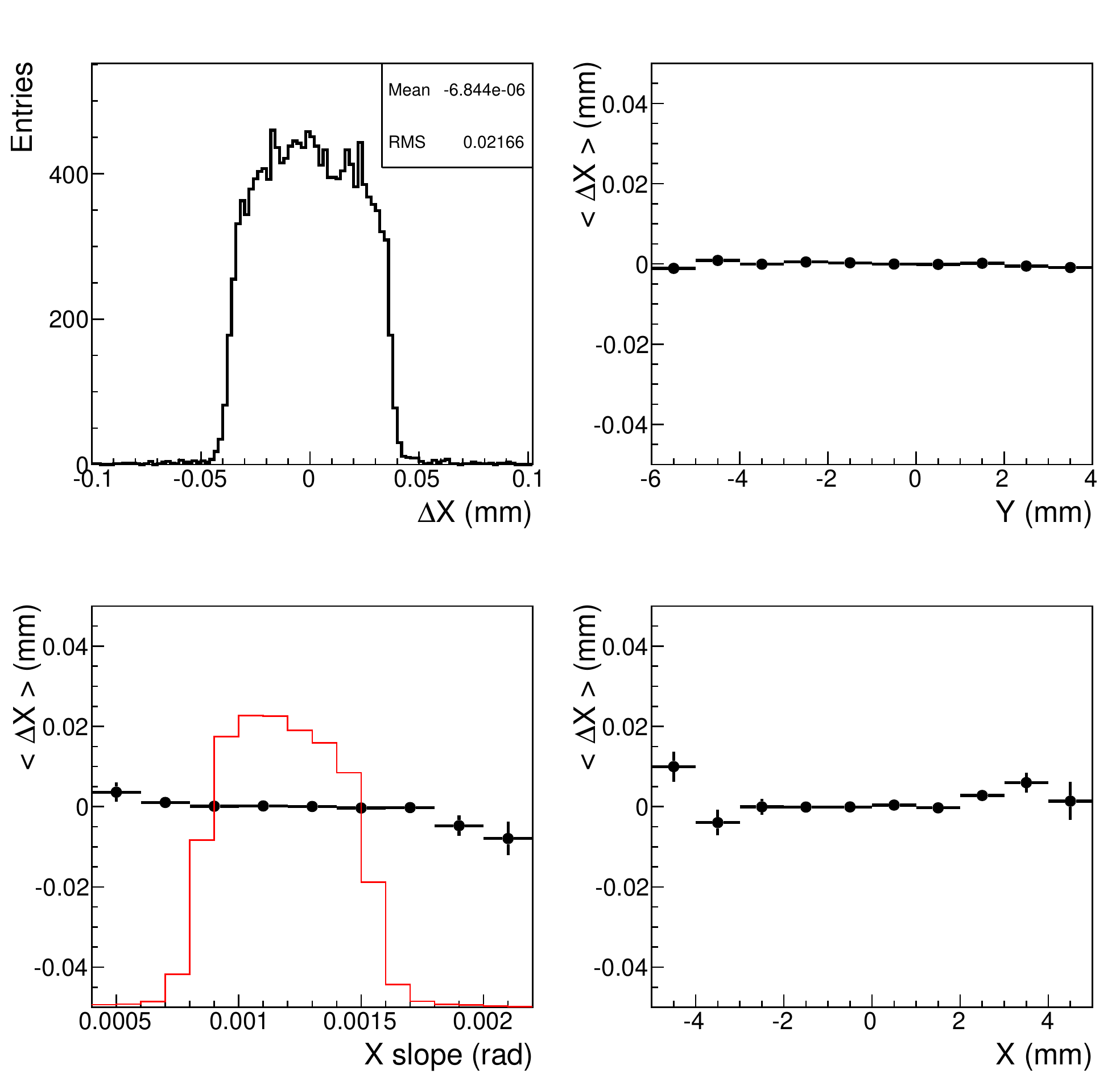}
\caption{\small{Alignment plots for sensor 3092-1-MS2, after alignment, showing (top left) $X$ residual, (top right) 
average $X$ residual vs $Y$ position, (bottom left) average $X$ residual vs $X$ slope, and (bottom right) 
average $X$ residual vs $X$ position, after the alignment is performed. The
$X$ and $Y$ positions and $X$ slope are obtained from the matching TimePix3 track.
Superimposed on the bottom left plot (red line) is the slope distribution of the tracks.}}
\label{fig:Align}
\end{figure}

The coordinate system is defined such that $X$ is horizontal, $Y$ is vertical, and $Z$ is parallel to the beam axis. 
Sensors are aligned with respect to tracks using the {\it residual} distribution, which is defined as the difference
between the reconstructed $X$ position of the hit and the $X$ position of the track extrapolated to the position of the sensor.
The degrees of freedom considered are: (i) offsets in the $X$ direction, (ii) offsets along the $Z$ axis, 
(iii)  rotations around the $Z$ axis, and (iv)  rotations around the $Y$ axis. 
A typical set of alignment plots are shown in Fig.~\ref{fig:Align}, after the alignment
is done. The profile plots are close to flat at zero. Small deviations are seen in the edges of the bottom
pair of distributions, but is likely an artifact of limited numbers of tracks and a possible bias due to the
steep dropoff of the beam profile near the edges.

\section{Results}
\label{sec:results}

The results of the test beam are presented below, starting with the results obtained from particles striking the detector at 
normal incidence, and continuing with the dependence on the incidence angle.

\subsection{Cluster finding in the DUT}
Clusters in the DUT are built up by searching for a {\it seed strip} that has a collected charge more than 20 ADC counts.
Moving away from the seed strip, we add adjacent {\it side strips} having at least 11 ADC counts (about 3$\times\sigma_{\rm noise}$).
The cluster is terminated when a side strip charge is below 11 ADC threshold. Thus, by definition, a cluster has 
between 1 and 5 strips included; each strip has charge of at least 11 ADC counts, and there is at least one strip with at least 20 
ADC counts. Therefore the minimum charge of any cluster is 20 ADC counts. When there are multiple strips in the cluster,
the position is computed using linear charge weighting, namely $\bar{x} = \sum{x_i q_i} / \sum{q_i}$, where $x_i$ and $q_i$
are the positions and charges of the strips in the cluster. Better resolution can be achieved with a non-linear correction,
but optimizing the resolution on these pre-prototype sensors is not a central goal of this test beam study.

\subsection{Tracking information}
\begin{figure}[tb]
\centering
\includegraphics[width=1.0\textwidth]{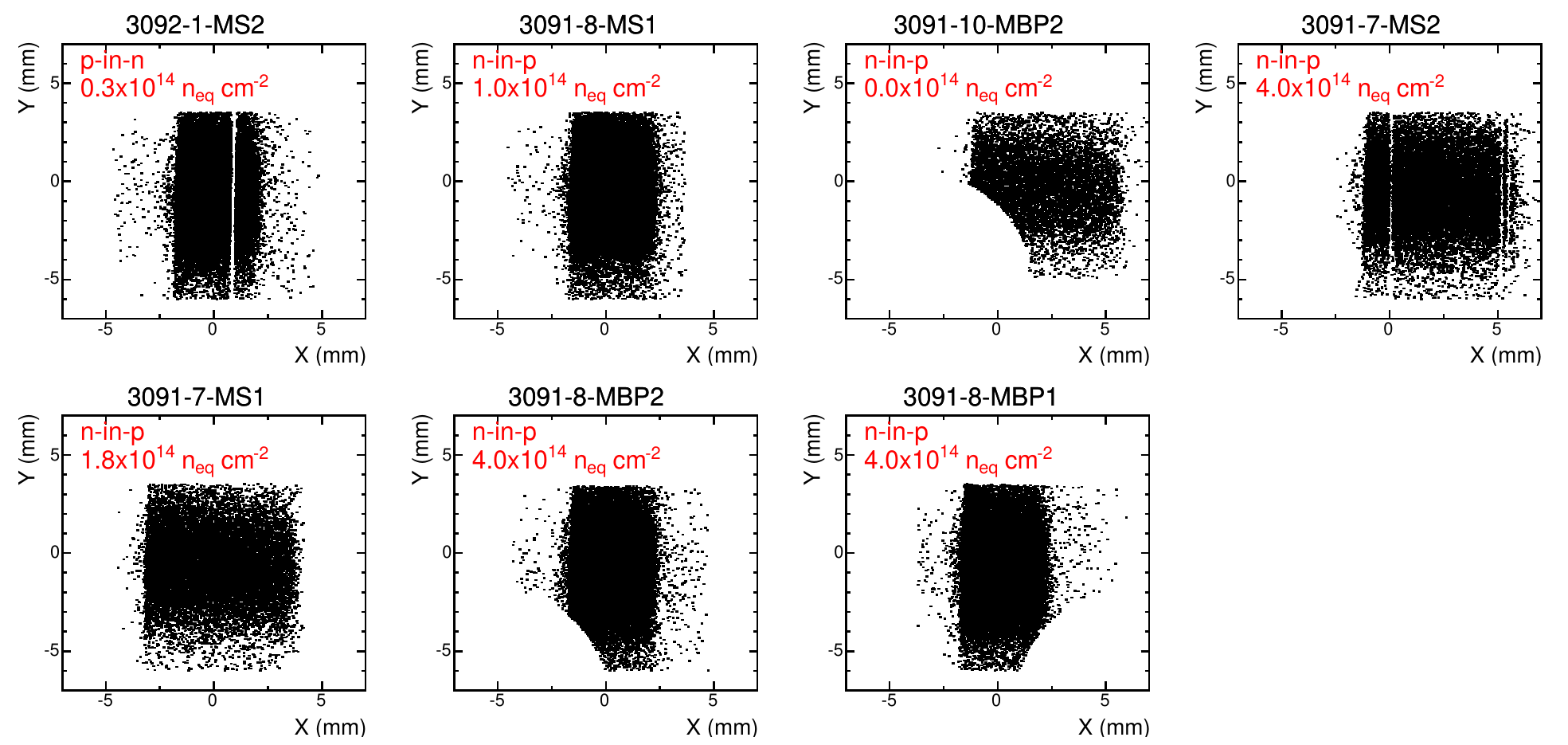}
\caption{\small{Profile of the beam ($Y$ vs $X$) at the DUT for each of the 7 sensors for a typical set of runs. The
3 DUTs with the quarter-circle edge are evident in these distributions.}}
\label{fig:YvsXTrack}
\end{figure}

Tracks were reconstructed using the TimePix3 telescope. About 95\% of tracks had 8-pixel hits; the remainder was 7-hit tracks.
Only tracks with good fit quality were used, by requiring that the $\chi^2$/ndf$<$4, where ndf is the number of degrees of freedom. 
Figure~\ref{fig:YvsXTrack} shows the beam profile ($Y$ vs $X$) for tracks that have a DUT hit within 100~$\mu$m from the track. 
The quarter-circle regions in board 2 chip 1, and board 4 chips 0 and 1 are evident. The sharp vertical edges are due to the collimators
in the beamline, and the boundaries along $Y$ are fiducial cuts used in the analysis.
The beam intensity was tuned so that most of the triggered events had a single beam track, although about 10\% of events had
more than one track. Multi-track events are excluded from the analysis, to ensure that the reconstructed track is the
one which produced the trigger.

\subsection{Single event displays}
\begin{figure}[tb]
\centering
\includegraphics[width=0.32\textwidth]{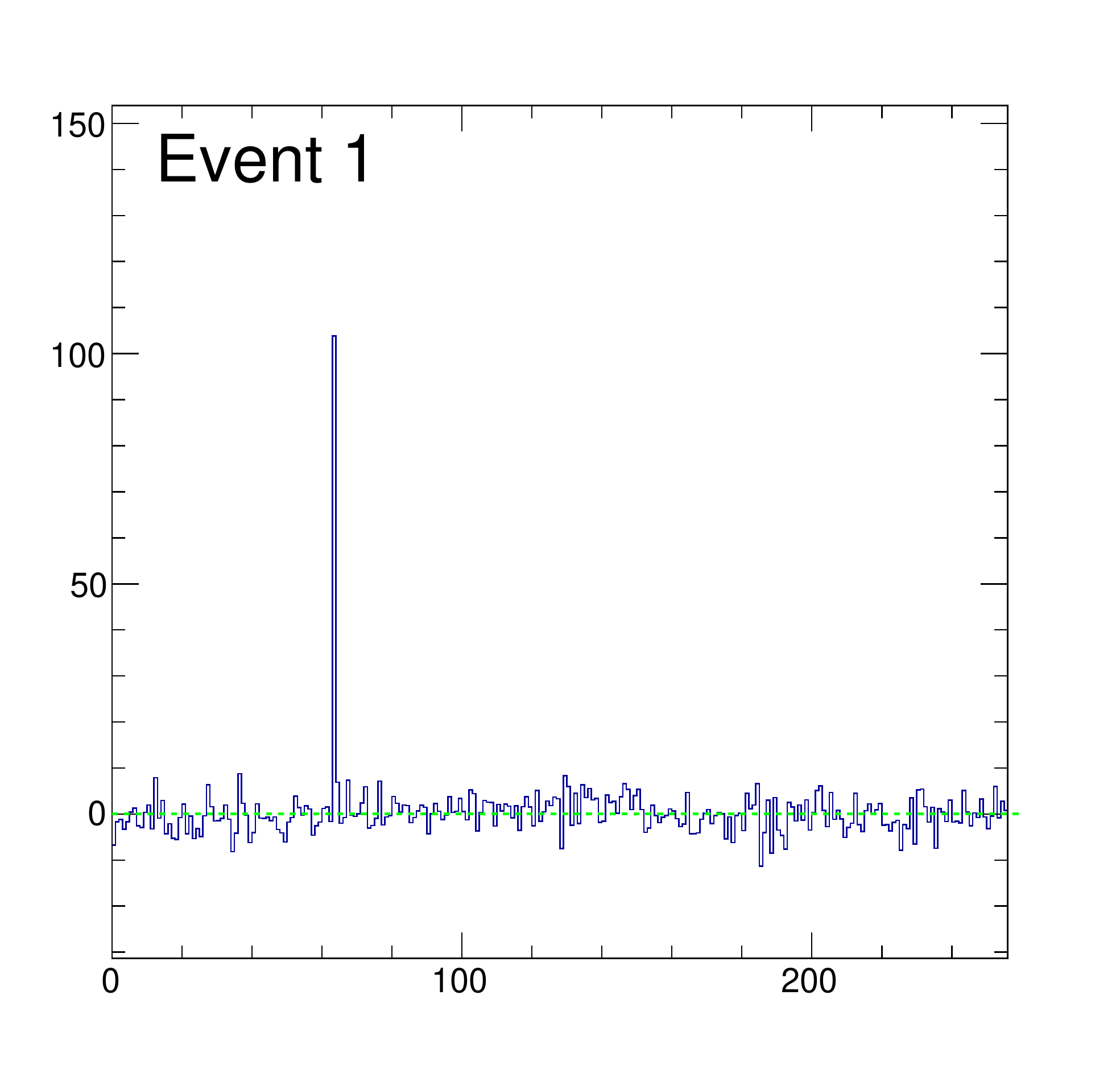}
\includegraphics[width=0.32\textwidth]{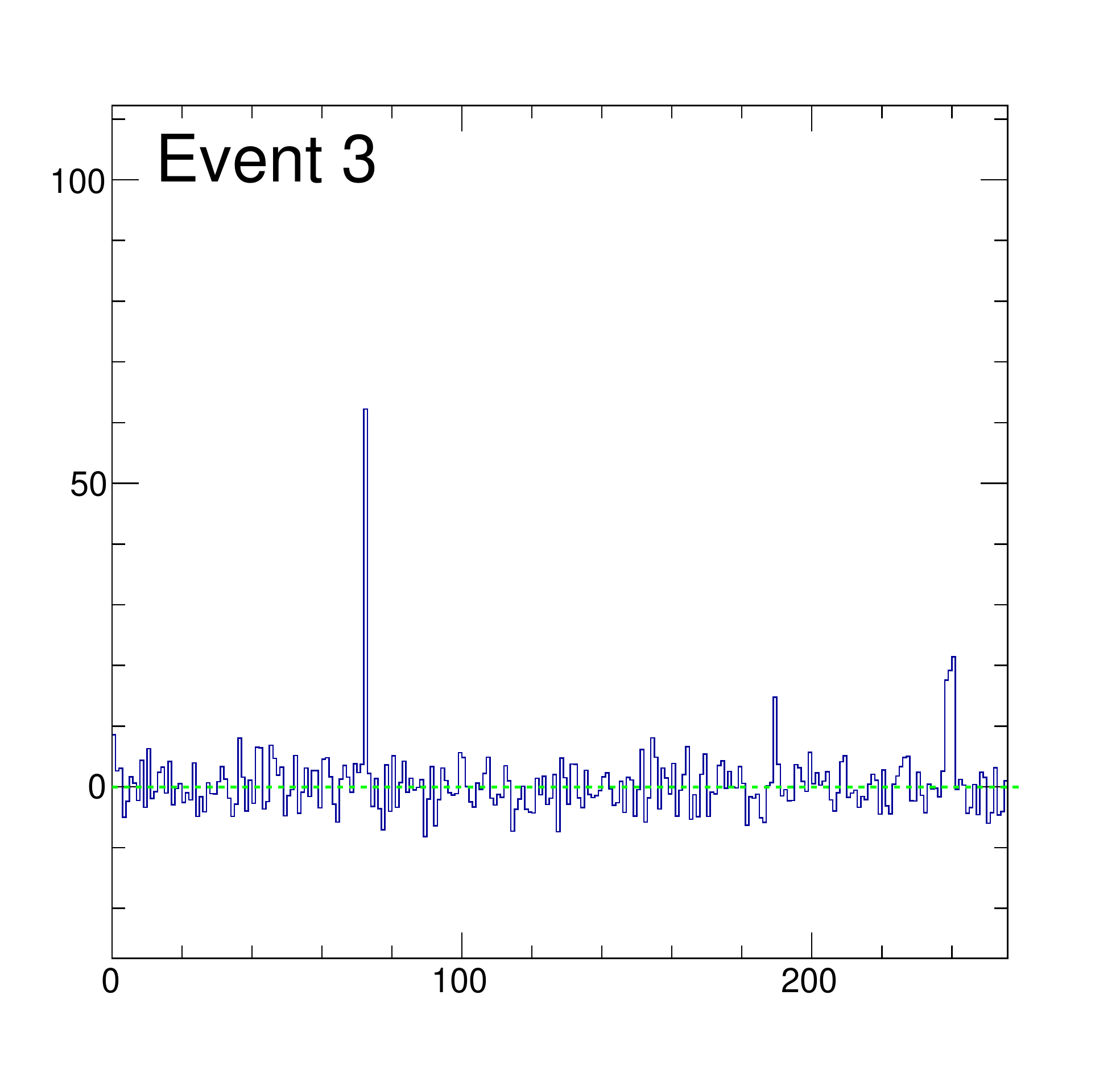}
\includegraphics[width=0.32\textwidth]{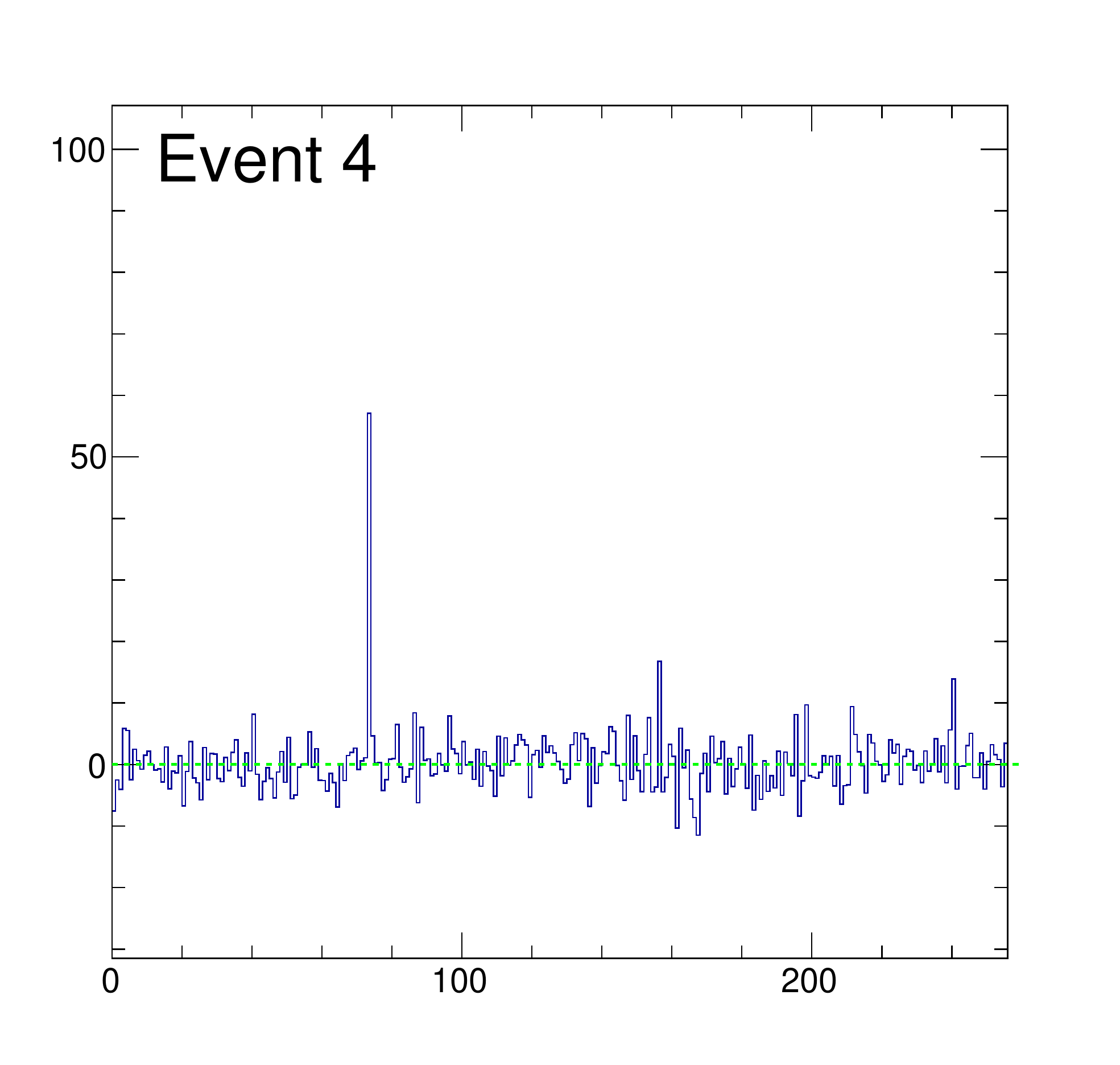}
\includegraphics[width=0.32\textwidth]{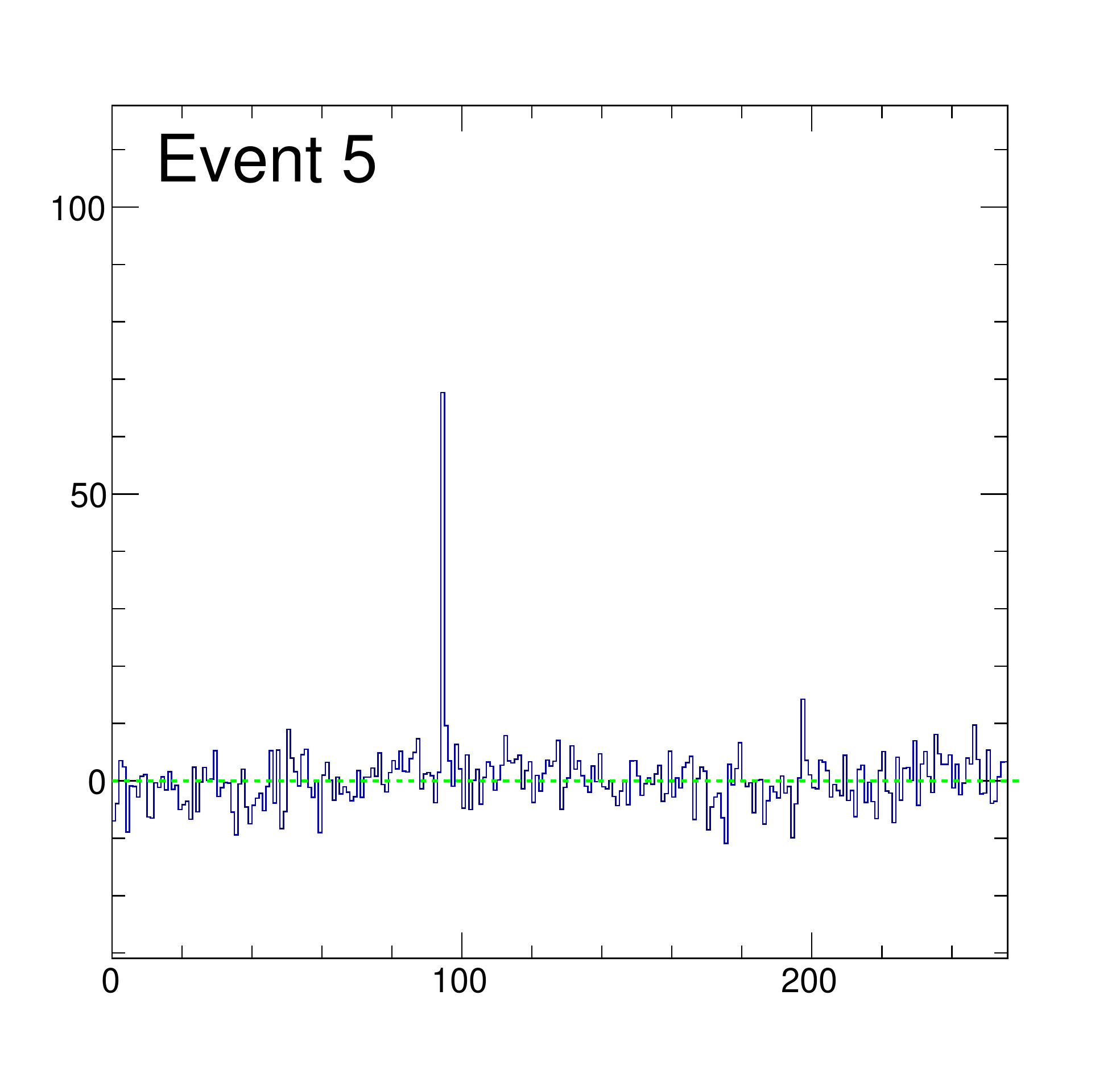}
\includegraphics[width=0.32\textwidth]{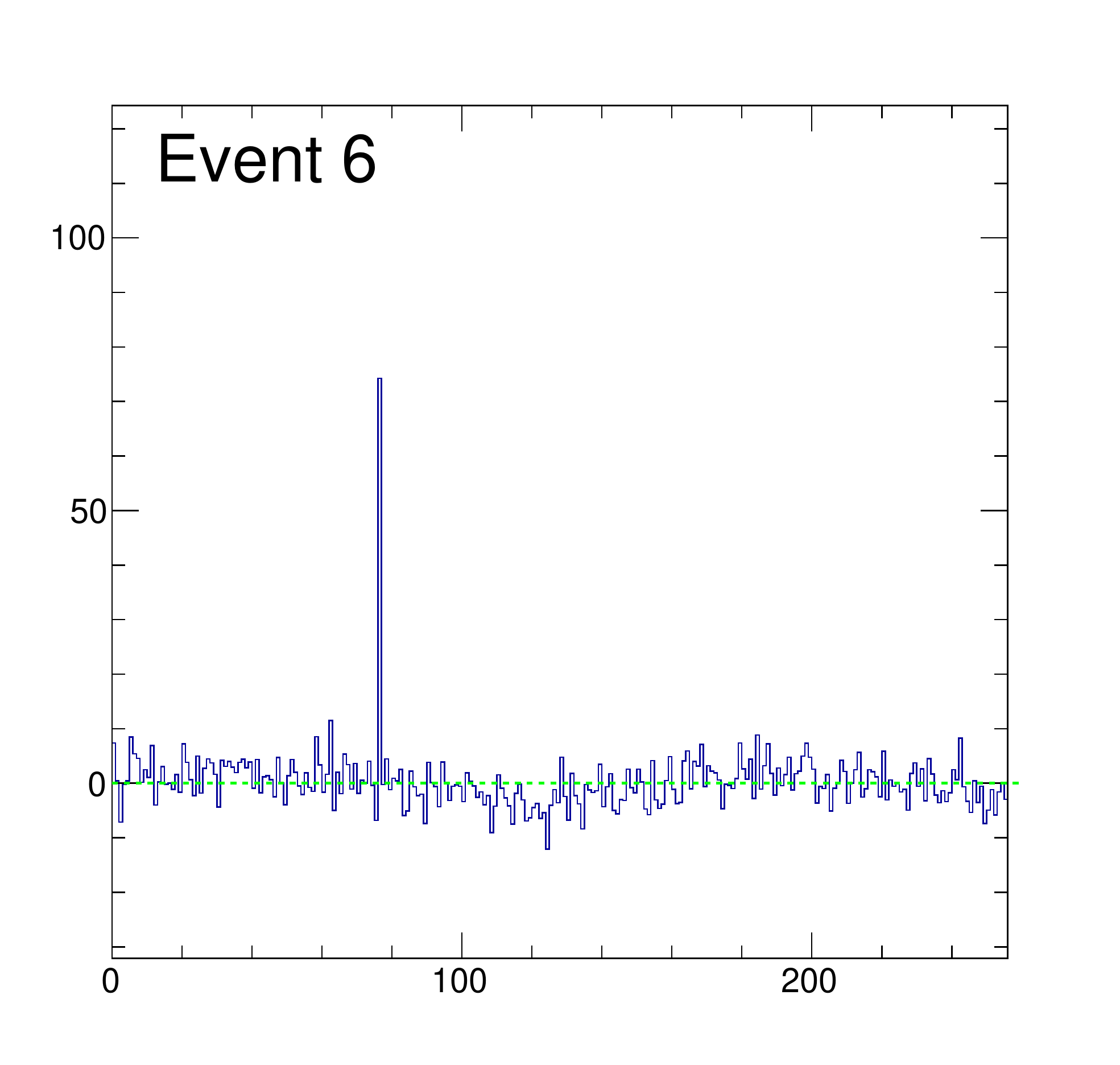}
\includegraphics[width=0.32\textwidth]{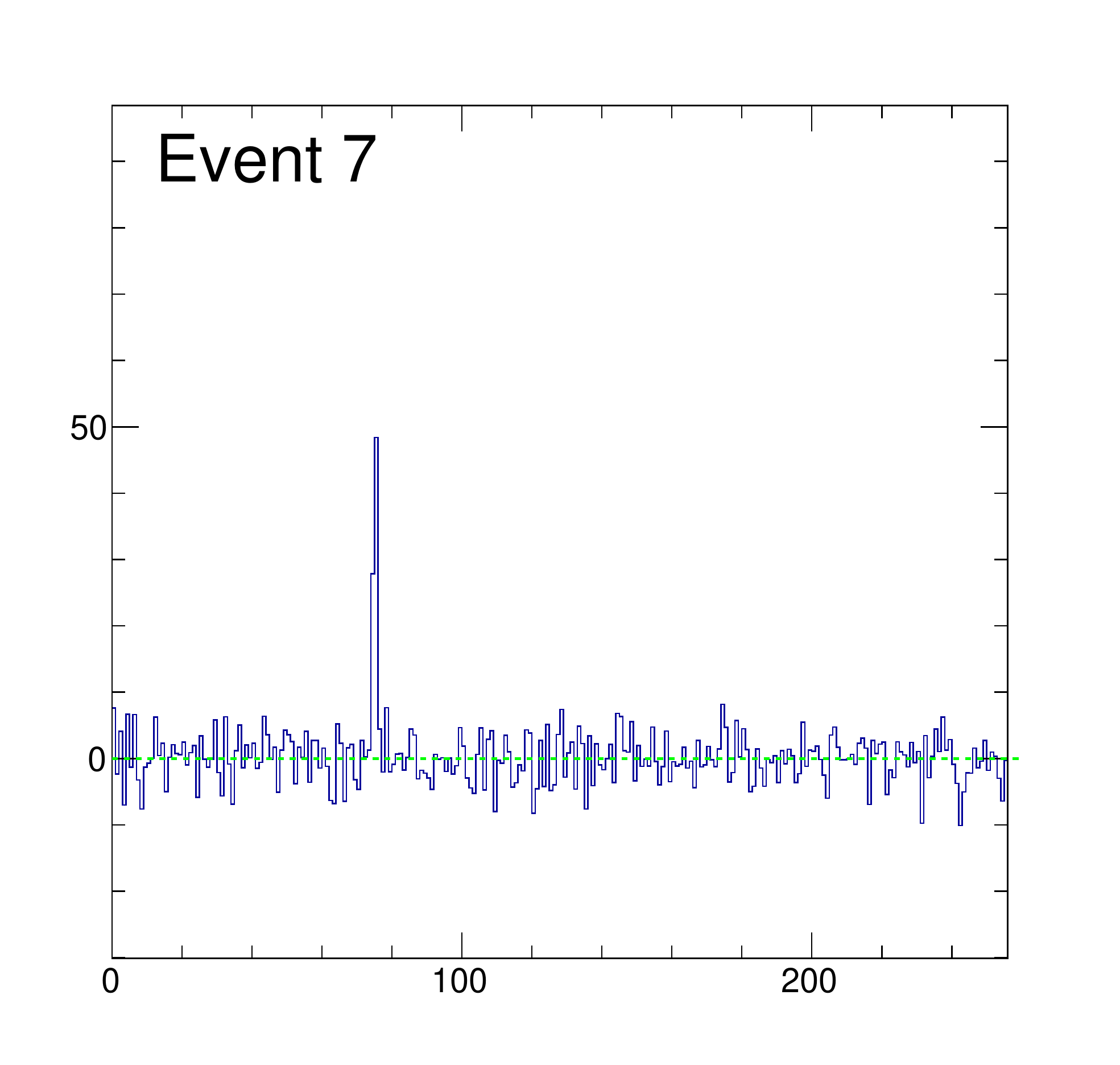}
\caption{\small{A few examples of single events for sensor 3092-1-MS2 (p-in-n), with a bias voltage of 250 V.}}
\label{fig:ExampleEvents}
\end{figure}

A few typical events, after all corrections are applied, are shown in Fig.~\ref{fig:ExampleEvents}. Here, no requirement is made on
the number of tracks. Strips with large ADC counts are indicative of the passage of a beam particle through the detector. 
The other channels show roughly Gaussian fluctuations about zero, typical of incoherent detector noise.
Signals generally stand out significantly above the noise.

\subsection{Noise}
\begin{figure}[tb]
\centering
\includegraphics[width=1.0\textwidth]{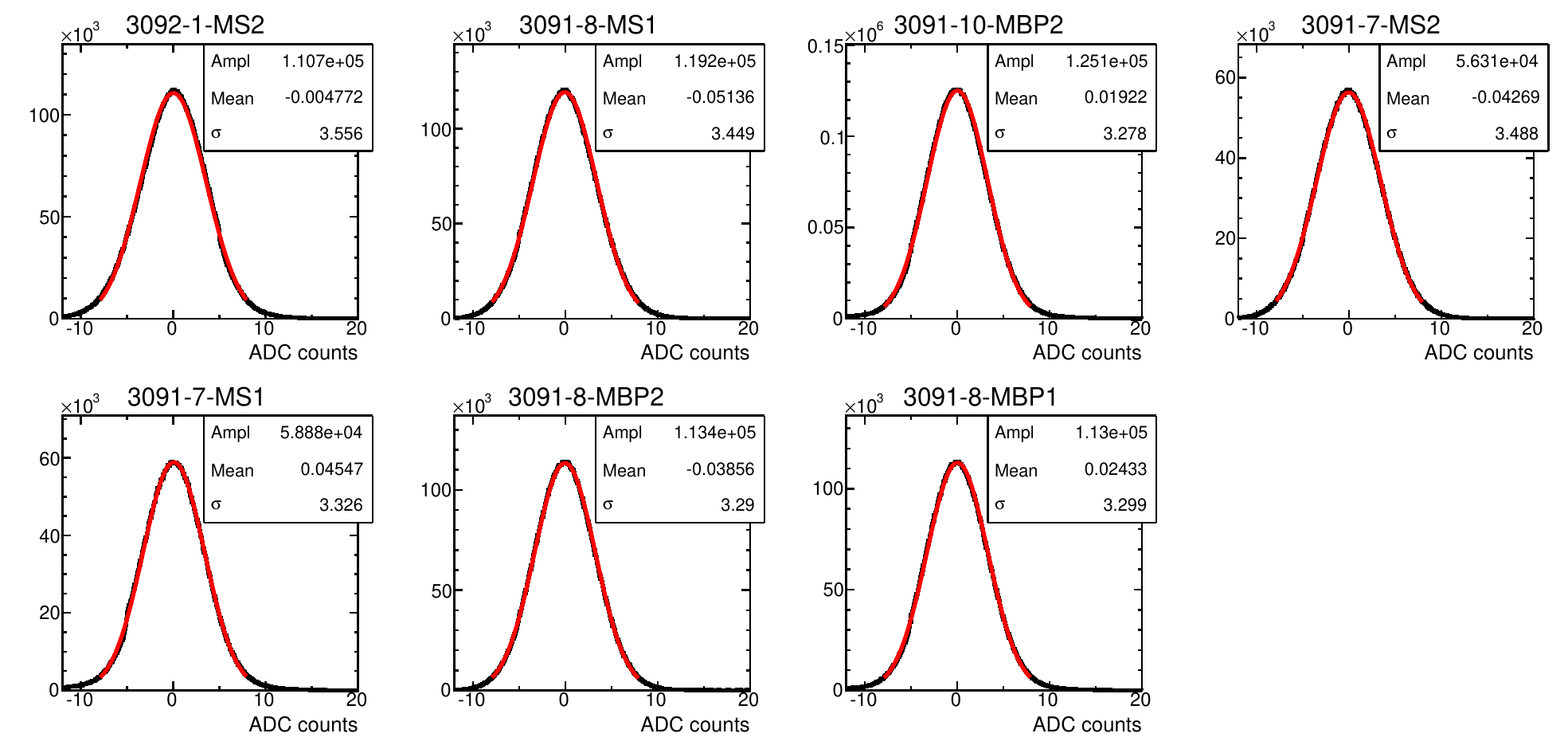}
\caption{\small{Noise distributions (in ADC counts) for the DUTs after all corrections during beam-on runs.
The red curves show a Gaussian fit to the core of the distribution.}}
\label{fig:Noise1D}
\end{figure}

To investigate the noise more quantitatively, the distribution
of ADC counts on all strips for a typical run of each sensor is shown in Fig.~\ref{fig:Noise1D}. The runs are chosen to be in a range where each
detector is fully depleted. While these are beam-on data,
the signal contribution to the shown distributions is negligible, since clusters typically have charge $\sim$70-80 ADC counts on average,
and are therefore off the plot. In addition, most events have only one cluster among the 255 channels.
For each distribution, the core of the distribution is fit to a Gaussian function and show the width, which is found to be in the range of 
about 3.3$-$3.5 ADC counts.
Since the charge calibrtation is about 275 $e^-$/ADC, this corresponds to a noise level of about 1000 $e^-$. Pedestal data with the beam off give compatible results.

\subsection{Landau distributions versus bias voltage}

\begin{figure}[tb]
\centering
\includegraphics[width=1.0\textwidth]{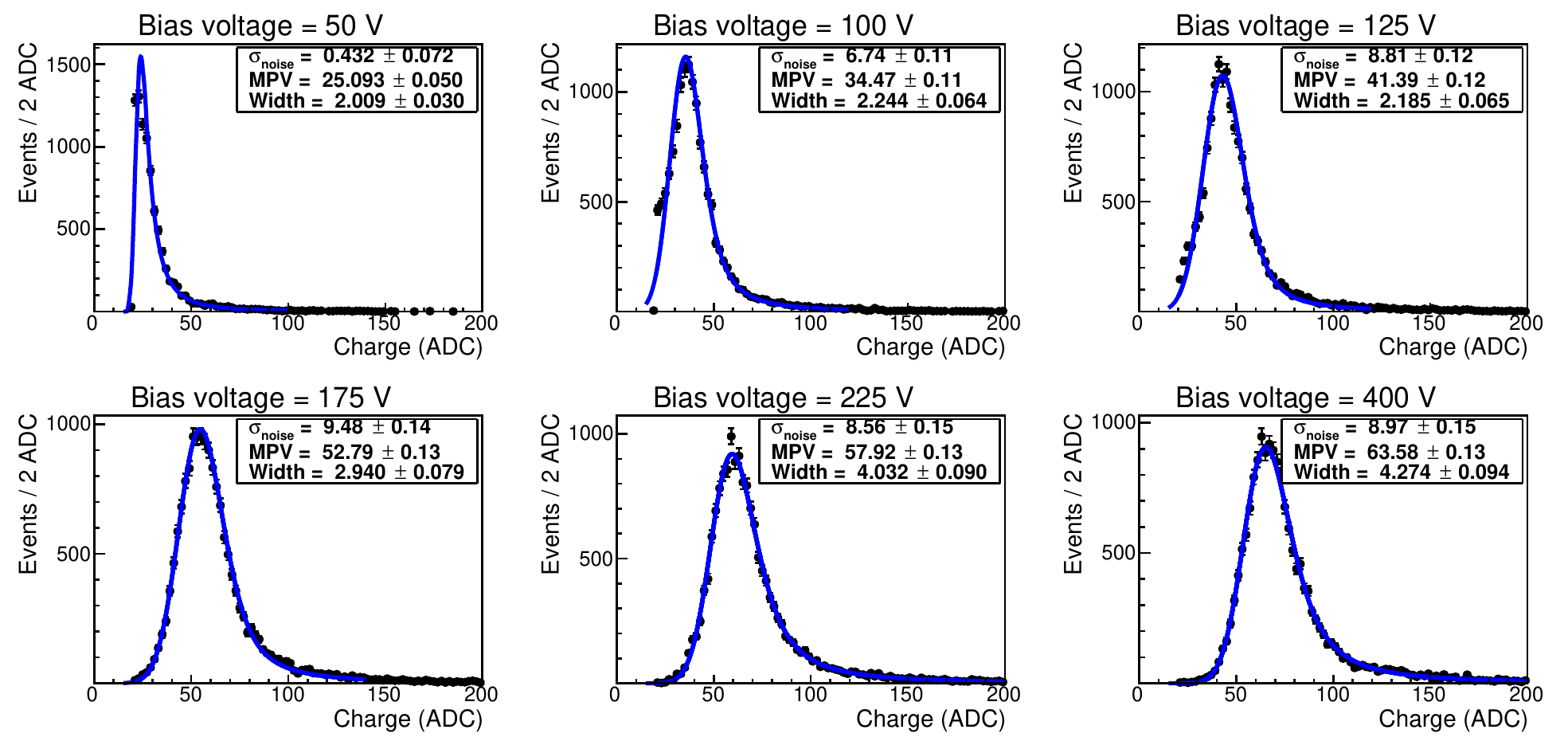}
\caption{\small{Signal distributions (in ADC counts) for sensor 3091-8-MBP2, which is an n-in-p type sensor, irradiated to $4.0\times10^{14}~n_{\rm eq}$/cm$^2$,
for several bias voltages. Each distribution is fitted to a Landau convoluted with a Gaussian resolution function and is overlaid.}}
\label{fig:Landau41}
\end{figure}

The most important aspect of these test beam studies is to validate that the sensors still have a high signal-to-noise ratio (S/N)
at the highest expected radiation fluence. For each sensor, data was recorded at bias voltages ranging from $50$~V (well below full depletion) to 
$500$~V (well above it). A few of the resulting distributions of cluster charge for sensor 3091-8-MBP2 (n-in-p, $4.0\times10^{14}~n_{\rm eq}$/cm$^2$)
are shown in Fig.~\ref{fig:Landau41} 
The cluster charge distributions are each fit to a Landau convoluted with a Gaussian function, and the fit parameters are indicated.
For bias voltages smaller than 75~V, the fits are biased because of the 20 ADC threshold requirement on the seed strip in the cluster. As 
the bias voltage increases, so does the collected charge.

\begin{figure}[tb]
\centering
\includegraphics[width=1.0\textwidth]{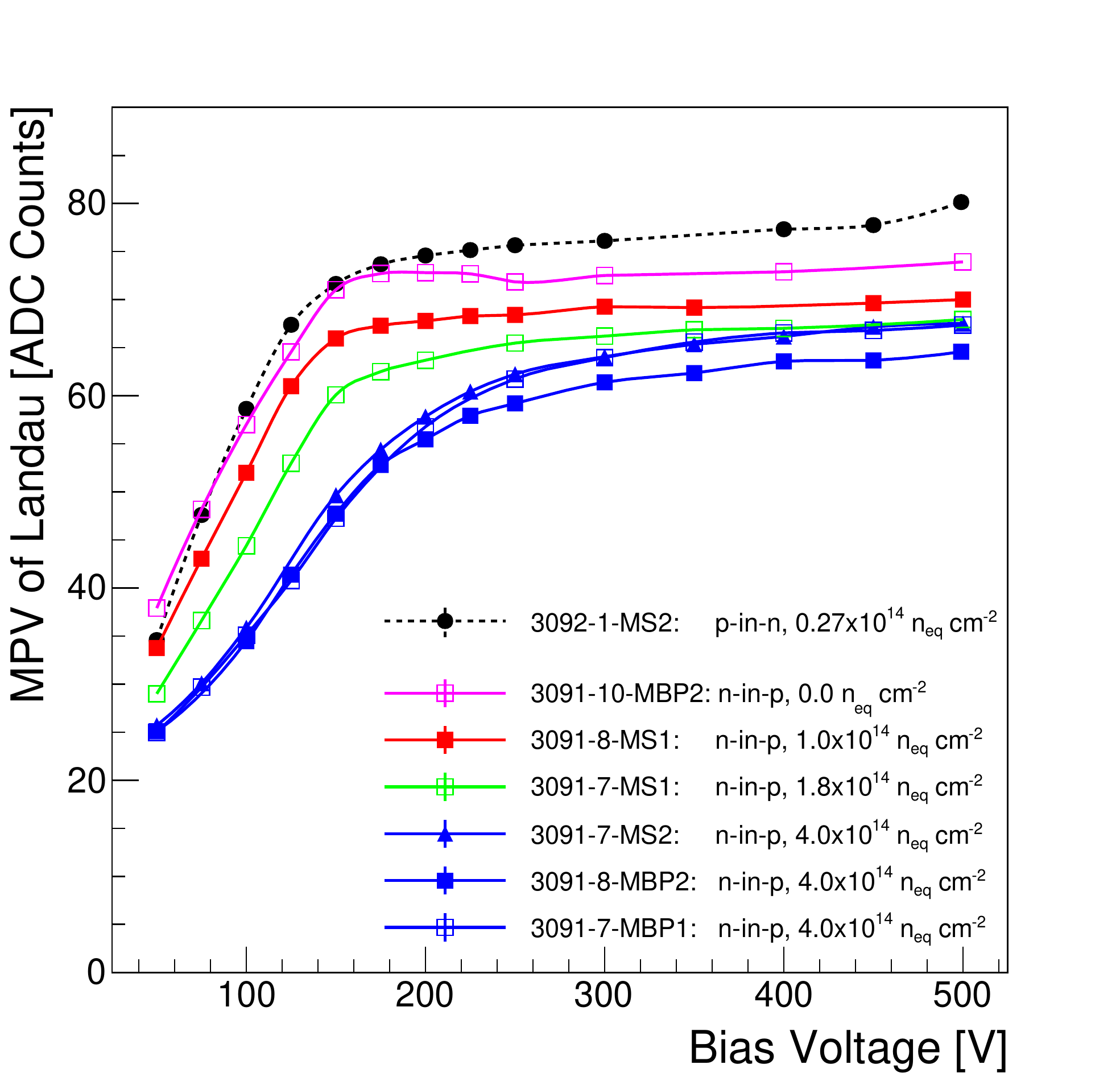}
\caption{\small{The most probable value of the collected charge for all cluster sizes
from the Landau fit as a function of bias voltage for each of the DUT. Tracks are at normal incidence to the sensor.}}
\label{fig:PlateauAll}
\end{figure}

The results for all DUTs tested are summarized in Fig.~\ref{fig:PlateauAll}, which shows the most probable value (MPV) of the Landau fit as a function
of the bias voltage, for each of the sensors. The p-in-n sensor is shown with a dashed curve and the n-in-p sensors are represented by 
solid curves. All of the curves exhibit a plateau in the $300 - 400$~V range, although there is a clear trend that the sensors with the
higher accumulated fluence exhibit a lower MPV in the plateau region. The loss is about 10\% at a fluence of 
$4\times10^{14}~n_{\rm eq}$/cm$^{-2}$, compared to no irradiation. Similar loss of collected
charge in highly irradiated detectors has been seen previously (see for example
Refs.~\cite{Casse:Vertex2007,Casse:Vertex2008,Artuso:2007bn,JWANG:Vertex2009,Affolder:2010zz}).
The three highest fluence ($4.0\times10^{14}~n_{\rm eq}$/cm$^2$) sensors show a similar behavior. A small difference in 
the MPV in the plateau is seen for one of the sensors, and this is most likely indicative of small differences in its 
charge calibration relative to the other two. Since the noise is $\sim$3.5 ADC per channel, the 
S/N is still quite high, about 18 for $V>400$~V, even in the most irradiated detectors. The S/N depends on
the detector capacitance, and hence we would expect smaller S/N in full size sensors.

\subsection{Cluster size}

\begin{figure}[tb]
\centering
\includegraphics[width=1.0\textwidth]{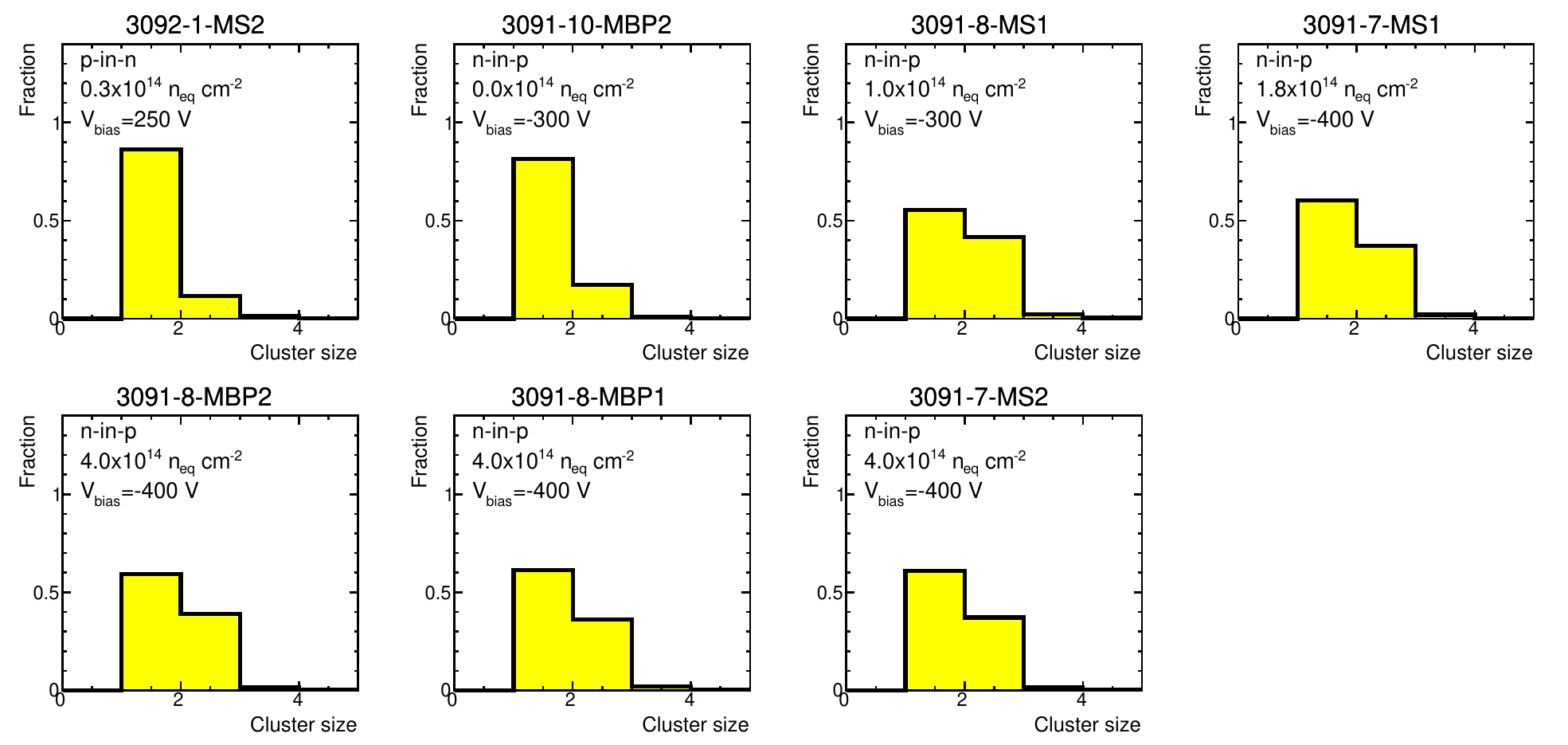}
\caption{\small{Distributions of the cluster size for the DUTs, after all corrections for tracks at normal incidence. The runs are taken from 
the plateau region of each DUT.}}
\label{fig:Cluster1D}
\end{figure}

Figure~\ref{fig:Cluster1D} shows the distribution of cluster sizes for the DUTs after all corrections are applied, all at 
normal incidence, and at bias voltages in the plateau region of Fig.~\ref{fig:PlateauAll}. In the p-in-n sensor, about 85\% of the 
clusters are single hit clusters. On the zero fluence n-in-p sensor, the single hit cluster fraction is a bit lower, about 82\%.
The irradiated n-in-p sensors show a substantially larger fraction of 2-strip clusters, of about 40\%. 
%The increase in cluster size can be understood as due to incomplete cancellation of the induced charges on
%neighboring strips due to charge trapping within the bulk and/or underdepletion of the detector~\cite{Brodbeck:1997tj,Borer:1999jk}.
While there is a loss of total charge collected in the highly irradiated detectors, a larger amount of 
charge sharing is observed.
%This effect, which leads to an increase in the cluster size, dominates over any decrease in cluster size due to the combination of 
%less of the total charge being collected and the fixed ADC thresholds used in the clustering.

\subsection{Cluster charge versus cluster size and interstrip position}
\begin{figure}[tb]
\centering
\includegraphics[width=1.0\textwidth]{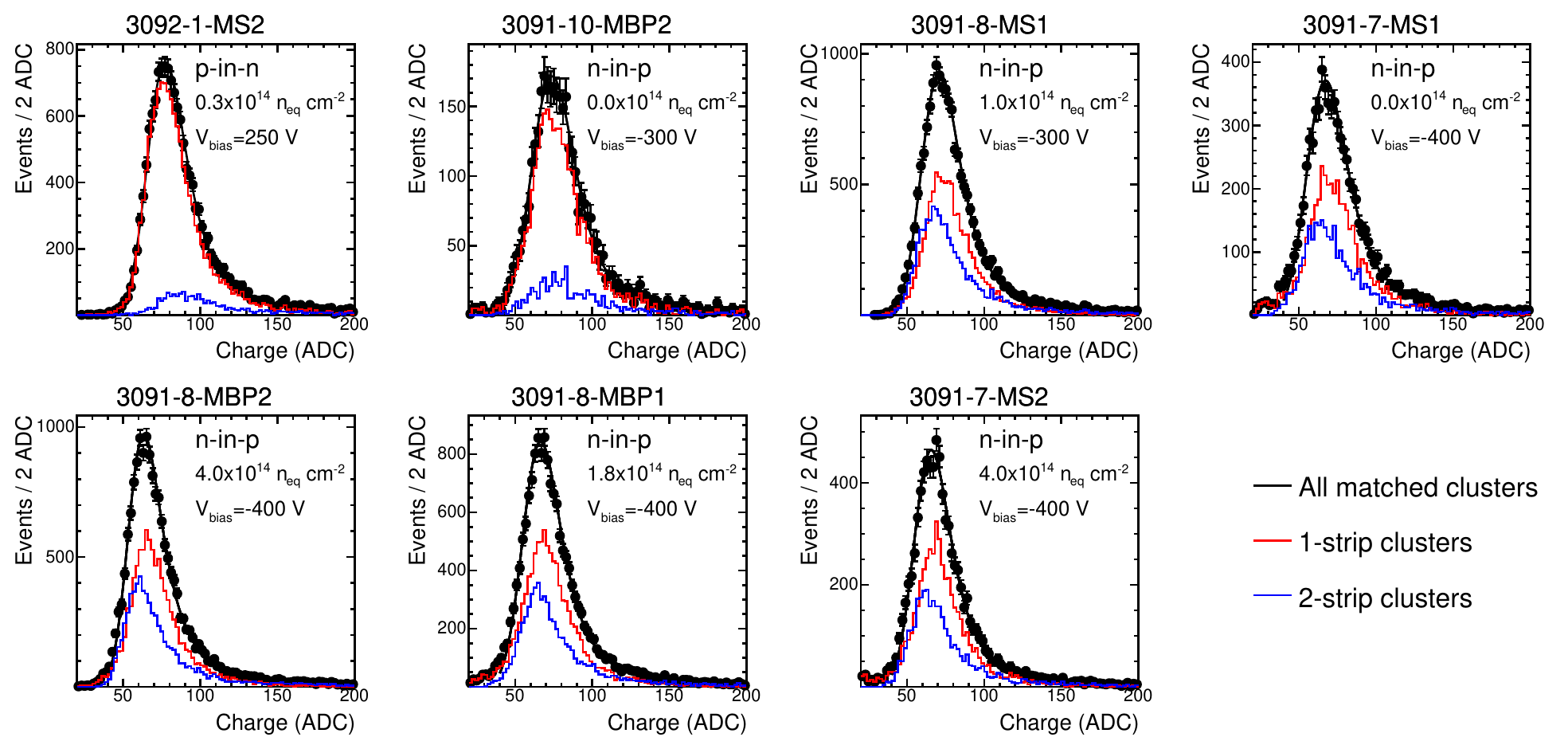}
\caption{\small{Distributions of charge collected in each DUT in the plateau region for tracks at normal incidence. The black points
show all clusters matched within 100~$\mu$m of a track, and the colored histograms show 1-strip and 2-strip cluster contributions.
The data are fit to a Landau convoluted with a Gaussian resolution function, and the fit is shown.}}
\label{fig:Landau1D}
\end{figure}

In this section, we investigate the dependence of the collected charge on the cluster size. With zero thresholds,
and no radiation effects, one would expect that the collected charge to be independent of the cluster size. However,
radiation effects and non-zero thresholds can lead to a difference. Figure~\ref{fig:Landau1D} shows the Landau distributions 
in the plateau region of the sensors, and overlaid are the
separate contributions from 1 and 2-strip clusters. The 1-strip and 2-strip clusters have similar MPVs, although
not identical. Due to the threshold bias mentioned above, the $0.27\times10^{14}~n_{\rm eq}$/cm$^2$ p-in-n and the zero fluence n-in-p sensors
have larger cluster charge for 2-strip clusters compared to the 1-strip clusters. However, for the irradiated sensors, the
opposite is seen, that is, 2-strip clusters have a lower charge collected than 1-strip clusters. 
From these observations, we conclude that the n-in-p sensors that received the radiation fluence (i) have reduced total charge,
(ii) have a larger fraction of two-strip clusters, and (iii) have less charge collected in two-strip clusters
compared to 1-strip clusters.

\begin{figure}[tb]
\centering
\includegraphics[width=1.0\textwidth]{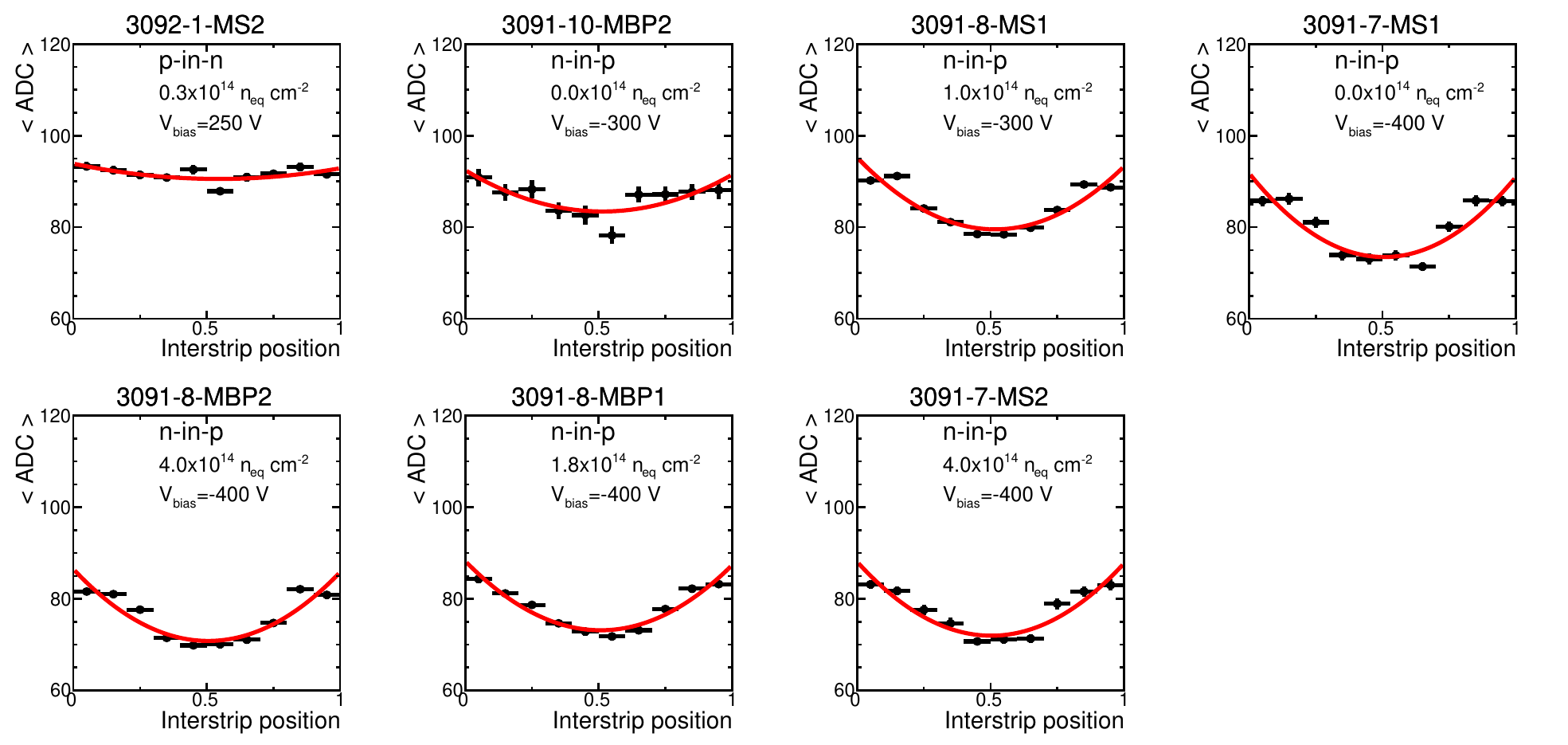}
\caption{\small{Distributions of the average cluster charge as a function of the relative position within a strip,
for tracks at normal incidence. A second order polynomial fit is overlaid just to guide the eye.}}
\label{fig:ChargeInterstrip}
\end{figure}

Additional information can be gained by investigating the charge collection as a function of the relative position
between two strips. For each track, the interstrip position is defined to be the
position at the DUT mapped onto the range from [0, 1], where 0 is the center of the $N^{th}$ strip and 1
is the center of the $(N+1)^{th}$ strip. The distributions of the average cluster charge as a function of the 
interstrip position are shown in Fig.~\ref{fig:ChargeInterstrip} for each of the DUTs.
It is seen that the heavily irradiated n-in-p sensors exhibit
a steady decrease in charge collection as the track approaches the middle region between two strips.
At the center, a loss of approximately 15\% in collected charge is observed.
The p-in-n and unirradiated n-in-p show a small loss in the middle between two strips, but
the effect is not more than a few percent. Similar loss of collected charge in the mid-region
between two strips in irradiated detectors has been reported in test beams of the ATLAS 
SCT~\cite{Campabadal:2005cn}.

From this, it is evident that a sizeable part of the drop in collected charge of the irradiated 
sensors is due to loss of charge when the track has an interstrip position between 0.2 and 0.8, where
a 10-15\% loss in charge is seen. When the track has an interstrip position less than 0.2 or greater than
0.8, the loss in collected charge is not as pronounced. Nevertheless, it is the average over the
strip which is relevant for the detector performance.
The efficiency versus the interstrip position has also been investigated, and in all cases, it 
is at least 99\%. Despite the loss in collected charge, there is no indication of any loss
of efficiency.

\subsection{Cluster size and resolution versus angle}

\begin{figure}[tb]
\centering
\includegraphics[width=0.48\textwidth]{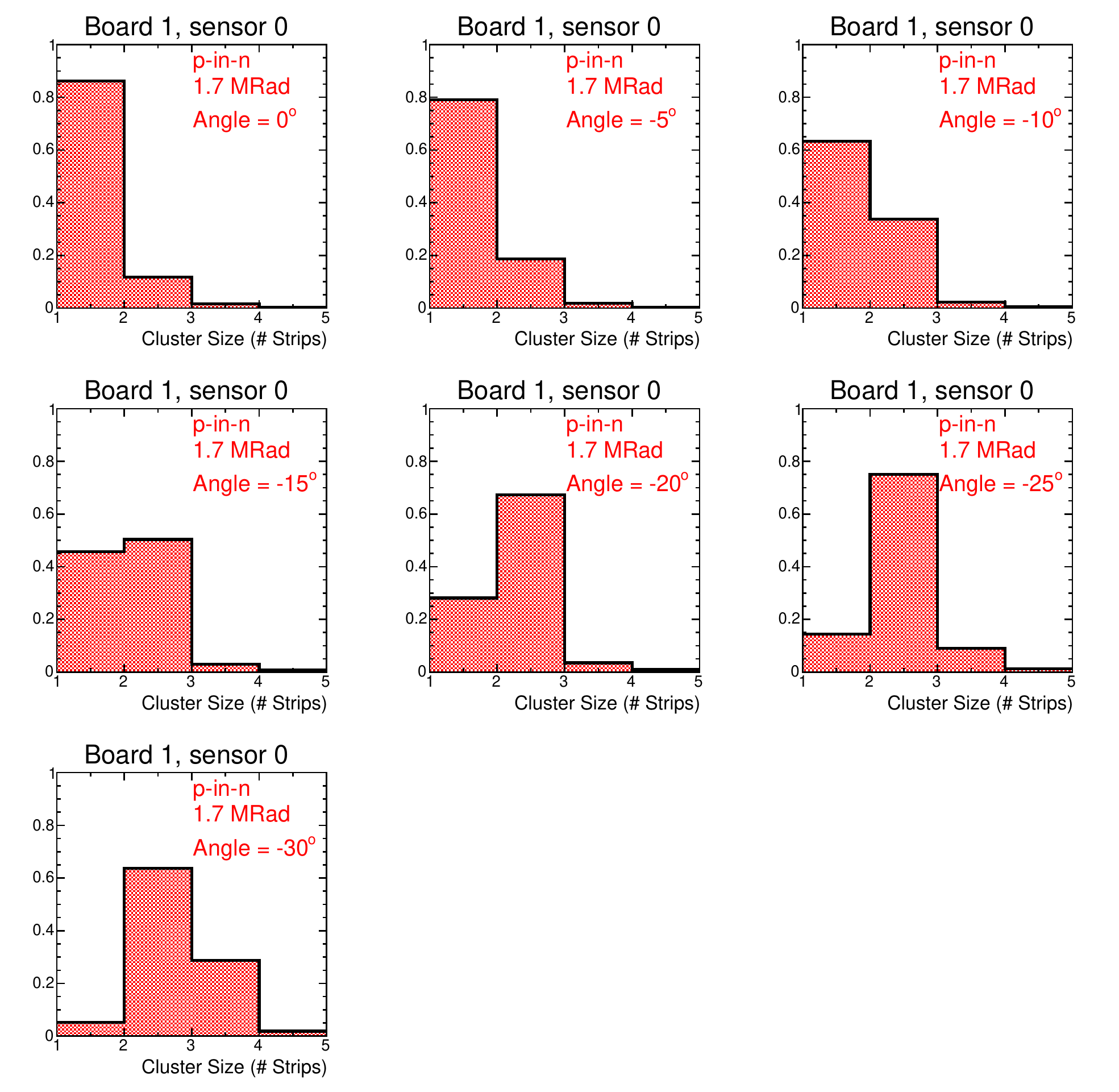}
\includegraphics[width=0.48\textwidth]{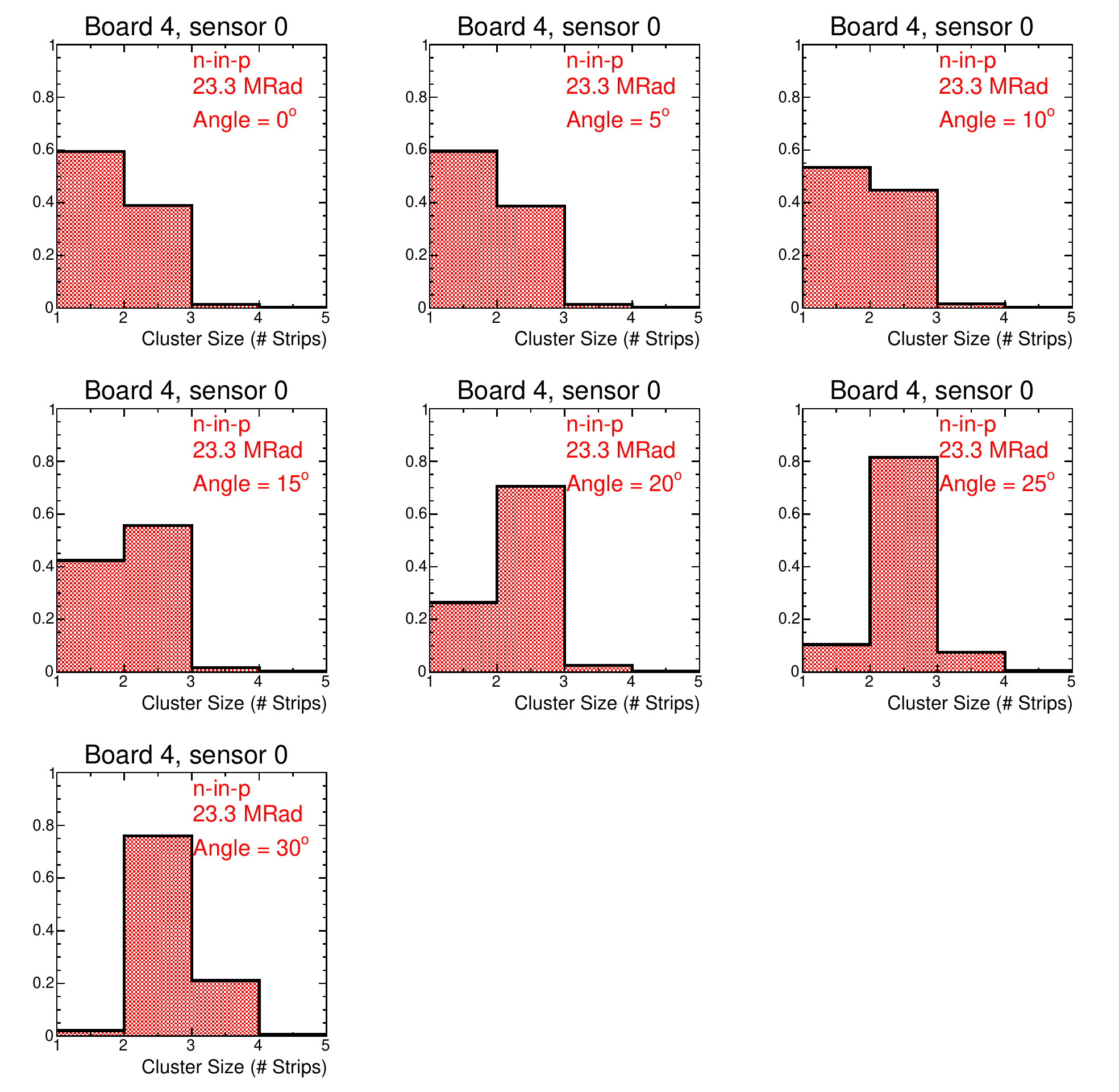}
\caption{\small{Fraction of 1-strip, 2-strip, and 3-strip clusters as a function of angle for (left) 3092-1-MS2 ($0.27\times10^{14}~n_{\rm eq}$/cm$^2$ p-in-n) at $V_{\rm bias}=250$~V, and (right) 3091-8-MBP2 ($4.0\times10^{14}~n_{\rm eq}$/cm$^2$ n-in-p) at $V_{\rm bias}=350$~V,.}}
\label{fig:ClusterSizeVsAngle}
\end{figure}

Studies of the detector performance were also carried out at angles ranging from normal incidence to 30$^{\rm o}$
(with respect to the normal to the sensor). The fraction of 1, 2 and 3-strip clusters as a function of the angle for 
(left) the p-in-n sensor and (right) one of the $4.0\times10^{14}~n_{\rm eq}$/cm$^2$ n-in-p sensors, is shown in Fig.~\ref{fig:ClusterSizeVsAngle}.
There is a clear increase in the fraction of 2-strip clusters in both cases, which reaches
a maximum at about 22.5$^{\rm o}$, beyond which the 3-strip clusters start to become sizeable. The main difference
between the two sensors is the cluster size at zero angle. For the heavily irradiated detectors, at normal incidence,
the 2-strip cluster fraction is significantly larger (see cluster size distributions for normal incidence in Fig.~\ref{fig:Cluster1D}).

\begin{figure}[tb]
\centering
\includegraphics[width=1.0\textwidth]{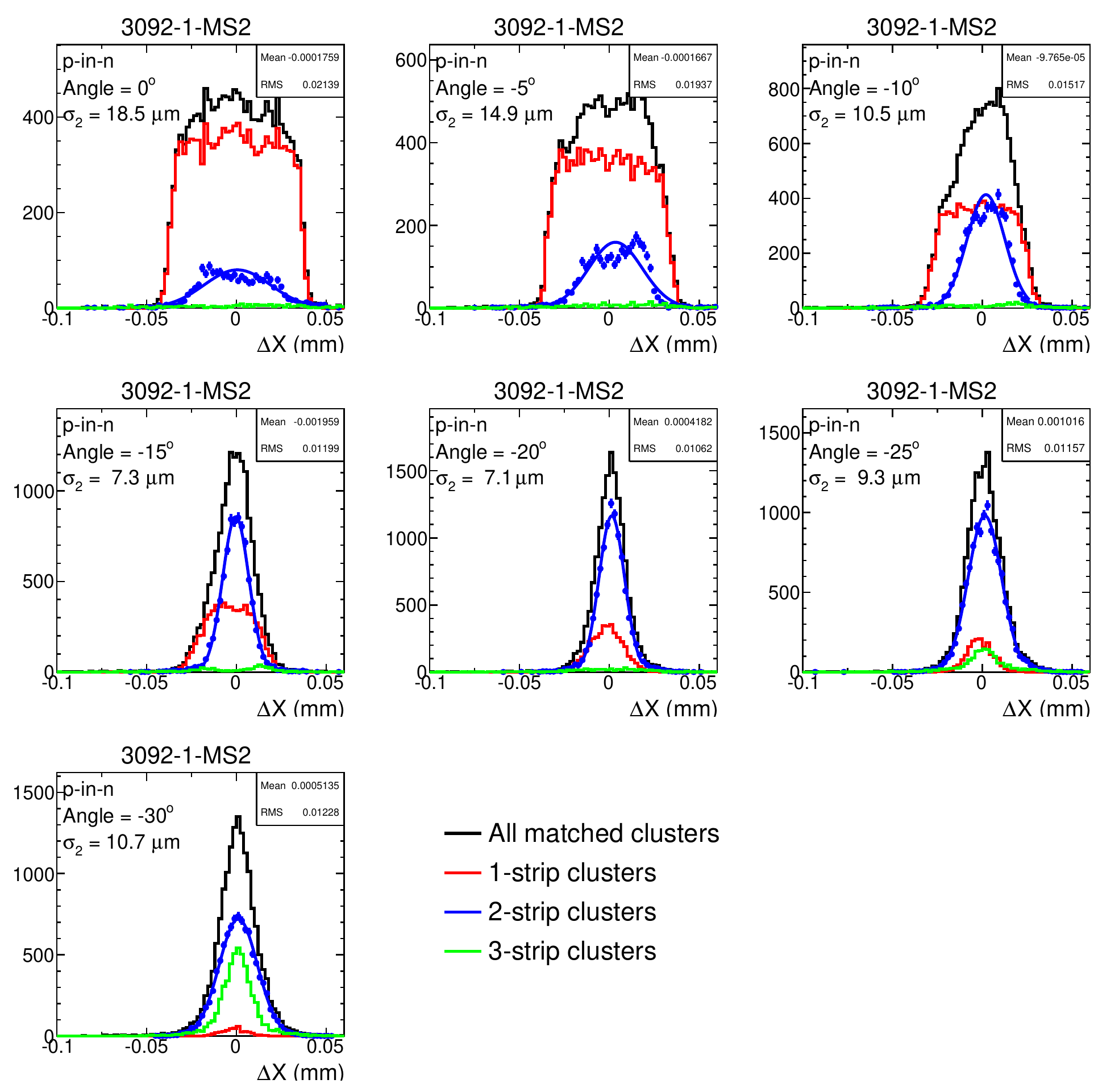}
\caption{\small{Residual distributions for angles ranging from $0^{\rm o}$ to $-30^{\rm o}$ for DUT 3092-1-MS2 ($0.27\times10^{14}~n_{\rm eq}$/cm$^2$ p-in-n) at $V_{\rm bias}=250$~V. 
The contributions from 1-strip, 2-strip, and 3-strip clusters
are also shown. The 2-strip cluster distributions are each fit to a single Gaussian function, and the widths, $\sigma_2$, are also indicated.}}
\label{fig:ResVsAngle_b1c0}
\end{figure}

The residual distributions between the UT hit and the track are shown in Fig.~\ref{fig:ResVsAngle_b1c0} for the $0.27\times10^{14}~n_{\rm eq}$/cm$^2$ p-in-n sensor.
The angles range from 0$^{\rm o}$ to 30$^{\rm o}$ in 5$^{\rm o}$ steps. The contributions from 1, 2, and 3-strip
clusters are also shown. The 3-strip clusters make up only a very small fraction, except at the two largest angles. 
As the angle increases the resolution improves, as indicated by the RMS of the distributions. The best resolution
should be achieved when $\tan^{-1}$(pitch/thickness)$\sim18^{\rm o}$.
The RMS is minimum for the $20^{\rm o}$ angle (RMS $=10.6~\mu$m.) The 2-strip residual distribution is fit to
a Gaussian function, and the width returned by the fit, $\sigma_2$, is shown. Two-strip clusters provide a resolution of 
about $7~\mu$m for tracks with angles in the 15-20$^{\rm o}$ range, 
which is about 3 times better than binary resolution (80\mum/$\sqrt{12}$). 
At small angles, most of the clusters are single strips, and the residual distribution
is roughly flat, as one would expect. At very large angles, the 2-strip resolution worsens. This is likely due to a third strip that should 
have been included in the cluster but was below the minimum ADC side strip threshold to be included. Improved resolution, primarily for
tracks at low incidence angle, was demonstrated by applying a non-linear charge weighting to two-strip clusters, but it is not
presented here.

\begin{figure}[tb]
\centering
\includegraphics[width=1.0\textwidth]{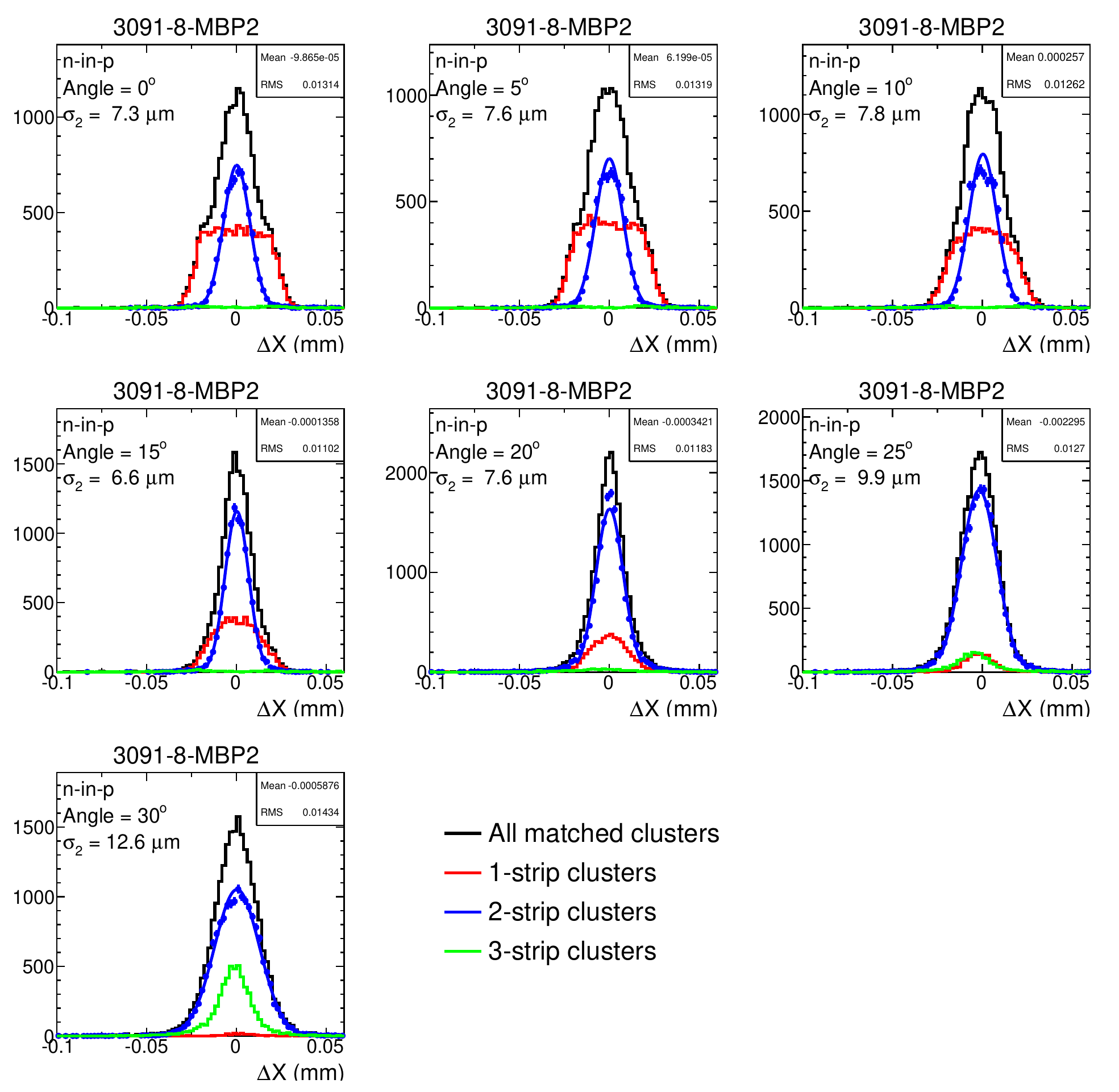}
\caption{\small{Residual distributions for angles ranging from $0^{\rm o}$ to $-30^{\rm o}$ for DUT 3091-8-MBP2 ($4.0\times10^{14}~n_{\rm eq}$/cm$^2$ n-in-p at $V_{\rm bias}=-350$~V).
The contributions from 1, 2, and 3-strip clusters are also shown. The 2-strip cluster distributions are each fit to a single Gaussian function, and 
the widths, $\sigma_2$, are also indicated.}}
\label{fig:ResVsAngle_b4c0}
\end{figure}

Figure~\ref{fig:ResVsAngle_b4c0} shows the corresponding residual distributions for one of the $4.0\times10^{14}~n_{\rm eq}$/cm$^2$ n-in-p sensors. A similar resolution
is obtained in the 15-20$^{\rm o}$ angular range, as for the p-in-n sensor. However, the notable difference (with respect to the p-in-n sensor) is that at low angles, 
the 2-strip contribution is sizeable, and shows the improved resolution that is generally accompanied by a 2-strip cluster. This clearly illustrates that the
larger cluster sizes associated with the irradiated n-in-p sensors are not an anomaly, since they are also accompanied by better position resolution.

\begin{figure}[tb]
\centering
\includegraphics[width=0.6\textwidth]{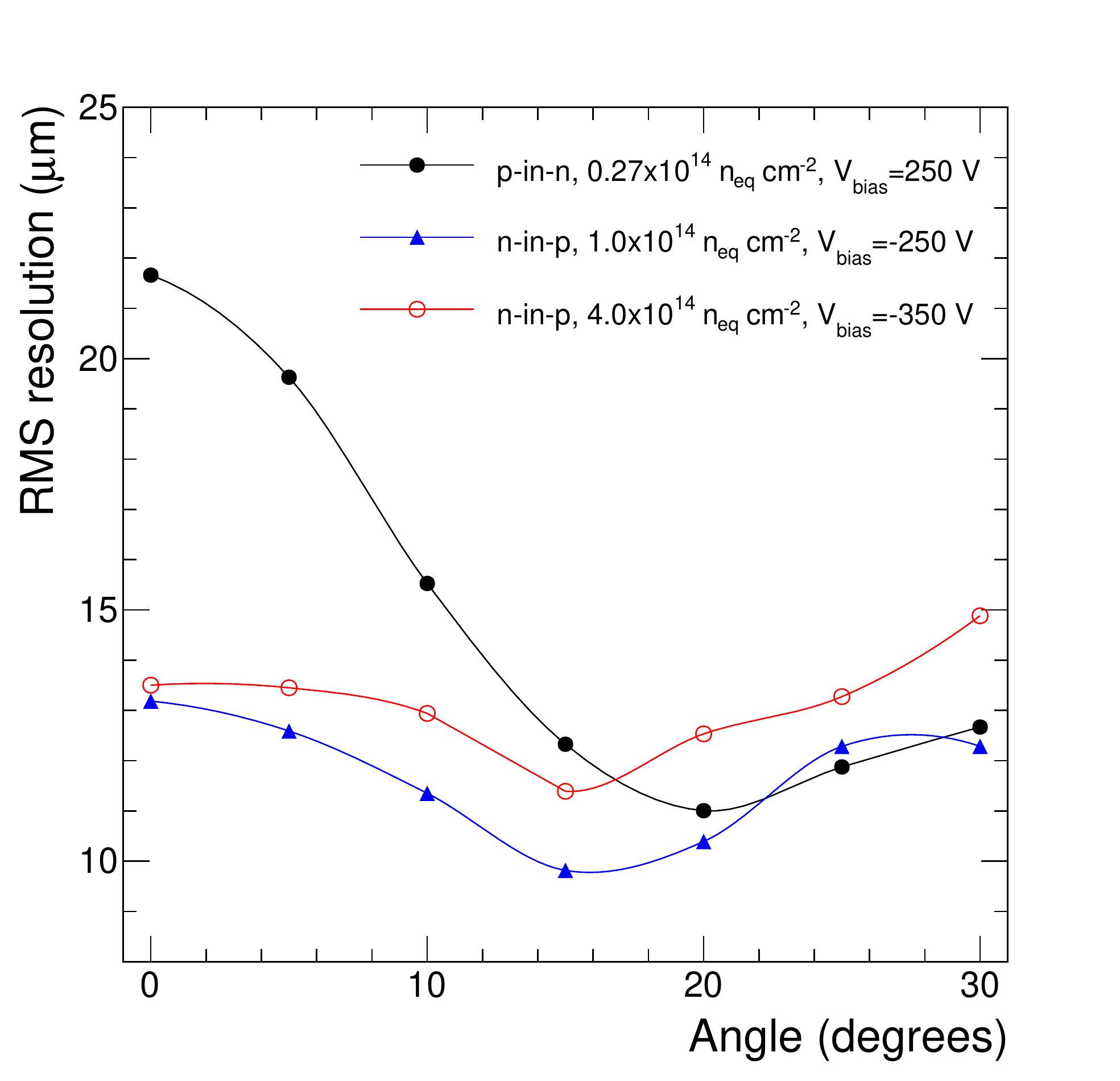}
\caption{\small{RMS of the residual distribution as a function of angle (with respect to the normal to the sensor) for 3 different sensors.
The bias voltages for the 3 sensors are 250~V for the}}
\label{fig:ResidualVsAngleGraph}
\end{figure}

The position resolution as a function of the angle is presented for 3 of the sensors in Fig.~\ref{fig:ResidualVsAngleGraph}, 
the $0.27\times10^{14}~n_{\rm eq}$/cm$^2$ p-in-n, the $1.0\times10^{14}~n_{\rm eq}$/cm$^2$ n-in-p and one of the $4.0\times10^{14}~n_{\rm eq}$/cm$^2$ n-in-p sensors. 
The resolution of the p-in-n sensor increases as one approaches 
normal incidence, reaching a maximum of
about $23~\mu$m, which is approximately binary, e.g. $80~\mu$m/$\sqrt{12}$. For the n-in-p sensors at normal incidence, the
position resolution is much better, about $13~\mu$m, owing to the larger amount of charge sharing between the strips (see Fig.~\ref{fig:ClusterSizeVsAngle}).
The $4.0\times10^{14}~n_{\rm eq}$/cm$^2$ irradiated n-in-p sensor shows slightly worse position resolution
than the $1.0\times10^{14}~n_{\rm eq}$/cm$^2$ sensor, but it is only at the $2~\mu$m level in the worst case. This difference is not of great concern for the UT detector.

\subsection{Charge collection near the quarter-circle region}

\begin{figure}[tb]
\centering
\includegraphics[width=0.9\textwidth]{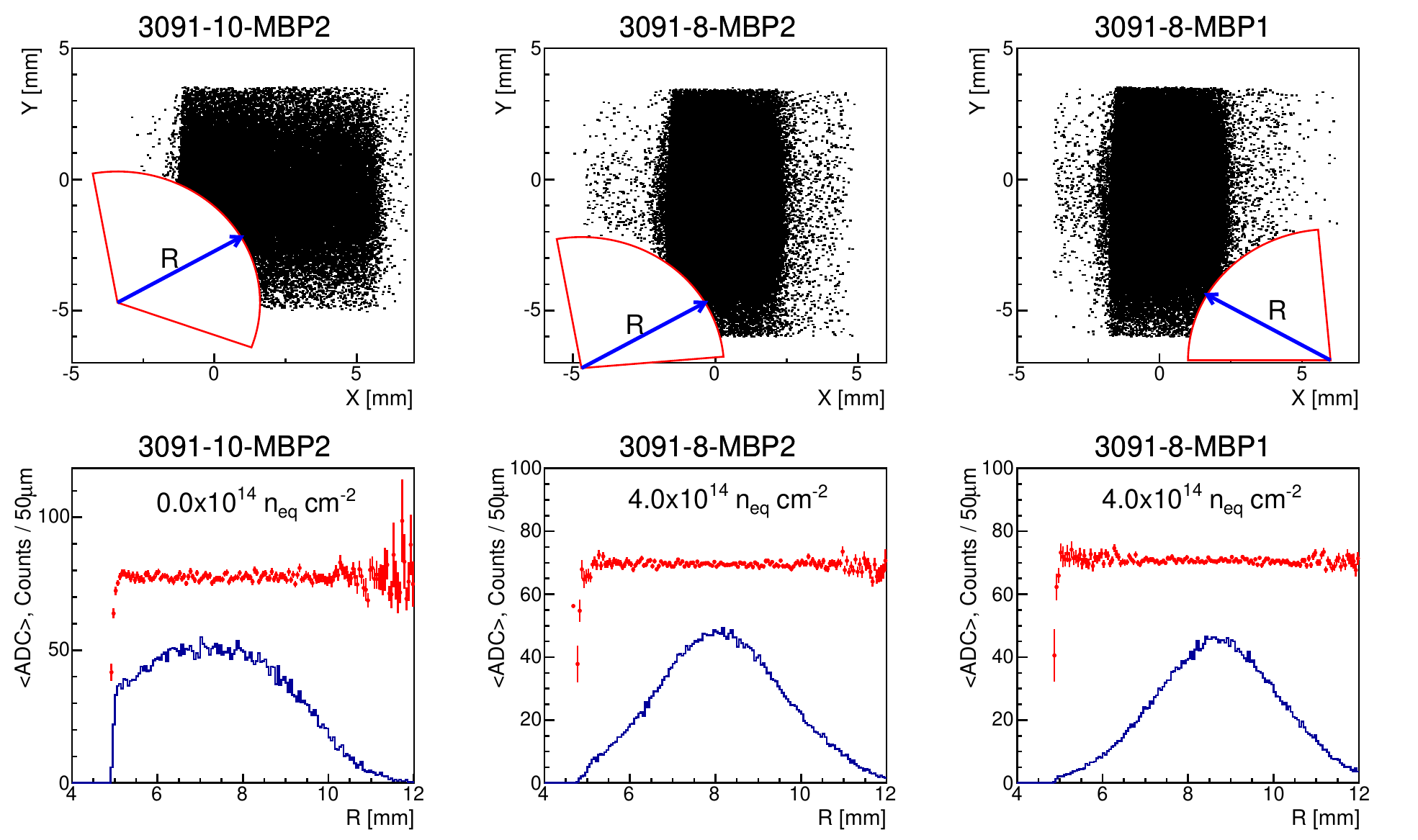}
\caption{\small{(Top row) XY distributions of tracks matched to clusters for each of the 3 mini beam-pipe sensors
for a bias voltage of -350~V.
(Bottom row) Overlay of average ADC value vs radial coordinate (red with error bars) and the radial distribution of 
tracks at the DUT (blue histogram). 
The radius is computed using the center of the red circles shown in the left column.}}
\label{fig:CutoutStudy}
\end{figure}

Three of the sensors studied during the test beam have a quarter-circle region where there are no strips 
(see Fig.~\ref{fig:sensors}). One of the sensors with this quarter-circle is the unirradiated n-in-p sensor,
and the other two are board 4 sensors 0 and 1. The latter two have different guard-ring structures, as
shown in Fig.~\ref{fig:sensors}. We investigate whether there is any drop in charge collected as one approaches
the quarter-circle radially. Using the data, the center of the circle is determined using the 3D information from
the TP3 tracks that have matched DUT clusters. The average ADC value as a function of the radial distance
from the center of the circle is then analyzed. The results for the 3 sensors are shown in Fig.~\ref{fig:CutoutStudy}, where each row
shows the results for a single sensor. The left column shows the XY profile of tracks that have matched DUT hits, along
with an arc that shows the estimated circle to match the quarter-circle region void of strips. The right hand plots
show an overlay of the radial distribution of tracks with matched DUT hits, with the average ADC value of clusters
as a function of the radius. In this figure, several runs with bias voltage above $350$~V are combined, and the average ADC is
the raw value with no TDC time requirement. This is done to increase the sample size, as the main interest is to
uncover a trend, while the absolute value is of less concern.

There is no clear indication of large loss of charge collection near the quarter-circle edge. 
A small loss in average charge in the final 50~$\mu$m or so from the edge cannot be ruled out, but the drop
is most likely just the precision on which the center of the circle is determined, which is about 50~$\mu$m. 
No significant difference between the two guard-ring structure designs (MBP1 versus MPB2 in bottom set of plots) is observed.
Moreover, no significant differences between the unirradiated and irradiated detectors is seen.

\section{Summary}
\label{sec:summary}

Results from a test beam at the CERN SPS of 7 pre-prototype UT sensors, one p-in-n and 6 n-in-p, have been presented.
The n-in-p sensors were irradiated at fluences of $1.0\times10^{14}~n_{\rm eq}$/cm$^2$, $1.8\times10^{14}~n_{\rm eq}$/cm$^2$
and $4.0\times10^{14}~n_{\rm eq}$/cm$^2$, and showed a gradual loss in total charge collected
with increased fluence. The maximum charge loss is about 10\% for the detectors with the largest
accumulated fluence. All detectors plateaued in the 300-400 V region, yielding a S/N of at least 15. 
With the full size UT sensors (10 cm x 10 cm, or 10 cm x 5 cm), there will be a larger input capacitance, 
and hence it is likely that the S/N will be lower. However, the studies presented indicate that
the anticipated 500~V maximum will be sufficient for operation of these detectors for the full 50~fb$^{-1}$ data set 
expected to be collected in Run III of LHCb.

We also find that the irradiated n-in-p DUTs have substantially larger cluster sizes at normal incidence,
compared to either the unirradiated n-in-p DUT, or the $0.27\times10^{14}~n_{\rm eq}$/cm$^2$  p-in-n DUT. These larger clusters sizes at normal
incidence are accompanied by improved position resolution, as one expects when there is charge sharing
between strips.

The dependence of the charge collection on the interstrip position has also been studied. For the irradiated n-in-p DUTs, it is found 
that about 15\% less charge is collected when a track goes through the middle of two strips, relative to the amount 
collected when it passes through the center of a given strip. This is consistent with findings from other experiments. 
The efficiency as function of interstrip position has also been studied, and in all cases it is found to exceed about 99\%.

The position resolution in the various sensors has also been investigated. At normal incidence, the irradiated detectors show
substantially better position resolution than the unirradiated detector. This stems from the larger fraction of 
2-strip clusters in the irradiated detectors. This goes against the conventional thinking that in irradiated detectors,
less charge is collected, resulting in a reduction of the fraction of multi-strip clusters.
At large angles, all detectors studied show improved spatial resolution, and the best resolution achieved 
is about 8 $\mu$m at an incidence angle in the $15-20^{\rm o}$ range. A resolution of about 6.5 $\mu$m is found
for the subset of 2-strip clusters.

The charge collection close to the circular portion of the DUTs was studied, and there is no indication of
any significant degradation of charge collection in the region close to this region. 
%Moreover, there is no clear difference between the two different guard-ring structures along this boundary.

% Do not include this in analysis note and conference reports
%\input{acknowledgements}
\section*{Acknowledgements}
 
\noindent We express our gratitude to our colleagues in the CERN
accelerator departments for the excellent performance of the SPS. 
INFN (Italy); MNiSW and NCN (Poland); SNSF and SER (Switzerland); and 
NSF (USA). We also acknowledge the efforts of the
VELO TimePix3 group, most notably Heinrich Schindler, Martin van Beuzekom, Hella Snoek, Paula Collins,
Raphael Dumps, Kazu Akiba, Xabier Cid Vidal, Alvaro Dosil, Bas van der Heijden
Sophie Richards, Dan Saunders, Panos Tsopelas, Jaap Velthuis and Mark Williams,
for providing a high precision telescope to enable many of the studies presented in this paper.
We also thank Ethan Cascio of the MGH proton irradiation facility for his support in carrying out
the sensor irradiations.
%\input{appendix}

% This should be taken out in the final paper
%\input{supplementary-app}

\addcontentsline{toc}{section}{References}
\setboolean{inbibliography}{true}
%\bibliographystyle{LHCb}
%\bibliography{main,LHCb-PAPER,LHCb-CONF,LHCb-DP,LHCb-TDR,REFS}

\ifx\mcitethebibliography\mciteundefinedmacro
\PackageError{LHCb.bst}{mciteplus.sty has not been loaded}
{This bibstyle requires the use of the mciteplus package.}\fi
\providecommand{\href}[2]{#2}

\end{document}